\documentclass[acmsmall]{acmart}
\usepackage{pifont}
\usepackage{menukeys} 
\usepackage{paralist}
\usepackage{amsmath}
\usepackage{epsfig,endnotes}
\usepackage{tikz}
\usepackage[ruled,vlined,linesnumbered]{algorithm2e}
\usepackage{amsfonts}
\usepackage{url}
\usepackage{array}
\usepackage{booktabs}
\usepackage{color}
\usepackage{xcolor}
\usepackage{pstricks}
\usetikzlibrary{calc,positioning,shapes.geometric,shapes.symbols,shadows,arrows}
\usepackage[justification=justified]{caption}
\usetikzlibrary{arrows}
\usetikzlibrary{patterns}
\usepackage{multicol}
\usepackage{lipsum}
\usepackage{adjustbox}
\usepackage{tabularx}
\usepackage{multirow}
\usepackage[english]{babel}
\usepackage{listings}
\usepackage{subcaption}
\usepackage{graphicx}
\usepackage{hyperref}
\usepackage{tcolorbox}
\usepackage{enumitem}
\usepackage{circledsteps}
\usepackage{makecell}
\usepackage{arydshln}
\usepackage{threeparttable}
\usepackage{wrapfig}
\usepackage{natbib}
\usepackage[normalem]{ulem}
\hypersetup{
    colorlinks = true,
    linkcolor = blue,
    anchorcolor = blue,
    citecolor = blue,
    filecolor = blue,
    urlcolor = blue
}
\definecolor{customred}{HTML}{FF5314}
\definecolor{customgreen}{HTML}{7CBB00}
\definecolor{mycolor}{RGB}{194, 214, 236}

\AtBeginDocument{%
  }

\setcopyright{acmlicensed}
\copyrightyear{2025}
\acmYear{2025}
\acmDOI{XXXXXXX.XXXXXXX}

\acmConference[xxx 2025]{xxx}{June xx--xx, 2025}{xxx, xx}

\acmISBN{978-1-4503-XXXX-X/18/06}

\newcommand{\sysname}{{\sc BinMetric}\xspace}
\begin{document}

\title{{\sysname}: A Comprehensive Binary Analysis Benchmark for Large Language Models}

\author{Xiuwei Shang}
\affiliation{%
  \institution{University of Science and Technology of China}
  \country{Hefei, China}
  }
\email{shangxw@mail.ustc.edu.cn}

\author{Guoqiang Chen}
\affiliation{%
  \institution{QI-ANXIN Technology Research Institute}
  \country{Beijing, China}
}
\email{guoqiangchen@qianxin.com}

\author{Shaoyin Cheng}
\authornote{Corresponding authors.}
\affiliation{%
  \institution{University of Science and Technology of China}
  \country{Hefei, China}
}
\email{sycheng@ustc.edu.cn}

\author{Benlong Wu}
\affiliation{%
  \institution{University of Science and Technology of China}
  \country{Hefei, China}
  }
\email{dizzylong@mail.ustc.edu.cn}

\author{Li Hu}
\affiliation{%
  \institution{University of Science and Technology of China}
  \country{Hefei, China}
}
\email{pdxbshx@mail.ustc.edu.cn}

\author{Gangyang Li}
\affiliation{%
  \institution{University of Science and Technology of China}
  \country{Hefei, China}
}
\email{ligangyang@mail.ustc.edu.cn}

\author{Weiming Zhang}
\affiliation{%
  \institution{University of Science and Technology of China}
  \country{Hefei, China}
}
\email{zhangwm@ustc.edu.cn}

\author{Nenghai Yu}
\affiliation{%
  \institution{University of Science and Technology of China}
  \country{Hefei, China}
}
\email{ynh@ustc.edu.cn}

\renewcommand{\shortauthors}{Shang et al.}

\begin{abstract}
Binary analysis remains pivotal in software security, offering insights into compiled programs without source code access. As large language models (LLMs) continue to excel in diverse language understanding and generation tasks, their potential in decoding complex binary data structures becomes evident. However, the lack of standardized benchmarks in this domain limits the assessment and comparison of LLM's capabilities in binary analysis and hinders the progress of research and practical applications.

To bridge this gap, we introduce \sysname, a comprehensive benchmark designed specifically to evaluate the performance of large language models on binary analysis tasks. \sysname comprises 1,000 questions derived from 20 real-world open-source projects across 6 practical binary analysis tasks, including decompilation, code summarization, assembly instruction generation, etc., which reflect actual reverse engineering scenarios. Our empirical study on this benchmark investigates the binary analysis capabilities of various state-of-the-art LLMs, revealing their strengths and limitations in this field. The findings indicate that while LLMs show strong potential, challenges still exist, particularly in the areas of precise binary lifting and assembly synthesis. In summary, \sysname makes a significant step forward in measuring the binary analysis capabilities of LLMs, establishing a new benchmark leaderboard, and our study provides valuable insights for the future development of these LLMs in software security. 


\end{abstract}


\maketitle


\section{Introduction}
Binary analysis is pivotal in various fields like software reverse engineering \cite{sutherland2006empirical}, malware prevention\cite{Hung2023SOICT}, and patch analysis \cite{xu2017spain}, serving a crucial role in understanding and dissecting the functionalities of software without access to its source code. According to a research report by Statista, approximately 21.5 billion IoT devices will be connected globally by 2025 \cite{Statista}. The diversity in instruction architectures and operating systems, coupled with the predominance of closed-source code and documentation, limits the applicability of source code analysis for securing IoT device firmware, which drives further updates of binary analysis technology.

Unfortunately, understanding and interpreting the structure and behavior of binary files is challenging due to their complexity and the lack of direct human readability \cite{Zhang2021sp,Patrick2020acsac}. Traditional tools \cite{IDA, Ghidra,BinaryNinja} and techniques \cite{ClangStaticAnalyzer}, while effective, often require extensive manual effort and expert knowledge, making the process time-consuming and frequently prone to errors. The integration of automated tools, especially those powered by artificial intelligence, has the potential to revolutionize this field by increasing efficiency and reducing human oversight.


In recent years, large language models (LLMs) have demonstrated increasing proficiency in a range of complex tasks \cite{Kang2023ICSE,chen2021evaluating,zhao2023survey,yu2024kieval,Lemieux2023icse,hou2024large,deng2023issta}, particularly showing promise in more specialized code-intensive areas such as code synthesis \cite{yan2023closer, HowEffective_ISSTA_2023, jiangICSE2023} and automated programming assistance \cite{wei2023copiloting, Leung2023ase, Pearce2022sp}. This has sparked questions among many software engineering practitioners: \emph{Can large language models like ChatGPT \cite{chatgpt} and CodeLlama \cite{codellama2023rozière} effectively perform binary analysis tasks?} However, the specific application of LLMs in the delicate area of binary analysis is still in its infancy, to some extent due to the lack of dedicated benchmarking frameworks that can adequately measure and drive progress in this area.

To address this limitation, we present \sysname, the first comprehensive benchmark designed to evaluate the capabilities of LLMs on binary analysis tasks, which supports multiple tasks following realistic reverse engineering scenarios. \sysname standardizes the evaluation process, providing a consistent and replicable framework for assessing the effectiveness of LLMs in this critical area. Specifically, \sysname is composed of six distinct tasks that mirror real-world binary analysis challenges, including call-site reconstruction, decompilation, signature recovery, binary code summarization, algorithm classification and assembly instruction generation. These tasks are built on 20 real open-source projects, ensuring the realisticity, diversity, quality, credibility, and maintenance of data sources. After data filtering and inspection, we extracted 1,000 question items from these projects. Furthermore, to evaluate these tasks from different dimensions, we built 4 evaluators and integrated them into an automated pipeline for easy one-click invocation. 

Next, to quantify the binary analysis capabilities of contemporary LLMs, we conduct an empirical study on \sysname to assess widely-used LLMs, including open-source models such as Llama2 \cite{Llama2023llama}, Codellama \cite{codellama2023rozière}, Mistral \cite{mixtral2024jiang}, WizardCoder \cite{wizardcoder2023luo}, DeepSeek \cite{bi2024deepseek}, DeepSeek-Coder \cite{deepseekcoder}, as well as closed-source like ChatGPT \cite{chatgpt} and GPT-4 \cite{openai2024gpt4}. The aim is to answer a couple of crucial questions: 

\vspace{-0.2ex}
\begin{center}
\begin{tcolorbox}[colback=gray!10,
                  colframe=black,
                  arc=1mm, auto outer arc,
                  width=13cm,
                  boxrule=0.75pt,
                  boxsep=-2pt
                 ]
\begin{itemize}[leftmargin=*]
    \item \textbf{RQ1}: What is the overall effectiveness of LLMs in binary analysis? 
    \item \textbf{RQ2}: Which LLM we investigated performs the best? And which type of LLMs performs better?
    \item \textbf{RQ3}: How efficient are LLMs? And what factors affect the effectiveness of LLMs?
\end{itemize}
\end{tcolorbox}
\end{center}
\vspace{-0.1ex}

Through the results of our empirical study, we obtain a set of findings. First, the LLMs demonstrate promising capabilities in binary analysis but struggle with tasks like call-site reconstruction and assembly instruction generation. Notably, each model exhibits expertise in a specific perspective. For instance, GPT-4 excels in binary lifting and logical analysis, while models like WizardCoder and CodeLlama perform well in semantic comprehension and assembly synthesis.
Second, GPT-4 leads in overall performance, and open-source models like CodeLlama-34B also show competitive capabilities, highlighting the potential of open-source solutions in this domain.
Finally, the efficiency of LLMs varies with model size and tasks. Larger models generally perform better but at the cost of efficiency. The design of one-shot prompts enhances effectiveness, while longer code inputs tend to reduce performance.

To summarize, our major contributions are in the following aspects:

\begin{itemize}
    \item \textbf{Benchmark.} We introduce \sysname, a pioneering comprehensive benchmark for assessing LLMs performance across multiple real-world binary analysis tasks. This benchmark includes 6 distinct tasks, 1,000 questions extracted and filtered from 20 real open-source projects, and 4 evaluators integrated into the automated evaluation pipeline.

    \item \textbf{Empirical Study.} We conduct the first large-scale investigation of widely-used LLMs using \sysname, studying (1) the overall effectiveness of LLMs across diverse binary analysis tasks, (2) the performance comparisons of different LLMs, and (3) efficiency of LLMs and factors affecting their effectiveness.


    \item \textbf{Findings and Insights.} Our results reveal the untapped potential of LLMs in binary analysis, providing new insights and future research directions for the field.
\end{itemize}


\vspace{-0.2ex}
\section{Background and Related Work}\label{sec:back}
\vspace{-0.5ex}
In this section, we first introduce our problem definition in \S\ref{sec:problem} and present the related works in \S\ref{sec:binanalysis} and \S\ref{sec:llm}.

\vspace{-0.5ex}
\subsection{Problem Definition}\label{sec:problem}
\vspace{-0.7ex}
Given a source code \emph{S}, it undergoes a compilation and stripping process to produce a binary file \emph{B}, represented as \emph{B} = \emph{R}(\emph{C}(\emph{S})), where \emph{C} denotes the compiler and \emph{R} denotes the stripping process of symbolic information. The binary code analyzer \emph{A}, designed to support a series of binary analysis tasks \emph{T} = \{\emph{t$_{1}$}, \emph{t$_{2}$}, \dots, \emph{t$_{n}$}\}, takes binary file \emph{B} as input and applies these tasks to generate corresponding outputs, formalized as:
\begin{equation}
\emph{O} = \emph{A}(\emph{B}) = \{\emph{o$_{1}$}, \emph{o$_{2}$}, \dots, \emph{o$_{n}$}\}.
\end{equation}
where each \emph{o$_{i}$} corresponds to the output of each task \emph{t$_{i}$}.

In this paper, we consider LLMs as binary code analyzers \emph{A}. Our objective is to rigorously evaluate their proficiency in executing these binary analysis tasks, ensuring that each output accurately reflects the intended analysis, maintaining a high level of fidelity and comprehensibility relative to the original source code \emph{S}, and show its effectiveness in various complex binary analysis scenarios.

\vspace{-1.0ex}
\subsection{Binary Analysis}\label{sec:binanalysis}
\vspace{-0.5ex}
Binary analysis involves examining and interpreting binary code, which is the machine-level representation of software executed by the CPU. Unlike source code, binary code is not human-readable and requires specialized techniques and tools to analyze \cite{david2020neural}. 
This process is crucial in various fields such as reverse engineering, malware analysis, and vulnerability discovery \cite{zhang2022pre}.

Traditional tools and techniques have been the backbone of binary analysis for decades. Disassembly tools, such as IDA Pro \cite{IDA} and Radare2 \cite{radare2}, convert binary code into assembly language, while decompilation tools like Ghidra \cite{Ghidra} attempt to revert binaries back to a more understandable, higher-level C-like pseudo code. Static analysis tools, such as Clang Static Analyzer \cite{ClangStaticAnalyzer}, examine binary code without execution, whereas dynamic analysis tools like Valgrind \cite{valgrind} and Pintools \cite{Pintools} track runtime behaviors in a controlled environment. Despite their widespread use, these traditional methods have significant shortcomings. They are often labor-intensive, requiring considerable expert knowledge and manual effort. Additionally, they struggle to extract high-level semantic information, which is crucial for understanding the broader context and functionality of the code. These methods are often inefficient when working with large codebases or highly optimized binaries, leading to incomplete or inaccurate analysis.

With the development of deep learning technology, many data-driven techniques have transformed the landscape of binary analysis. These methods leverage large datasets and advanced algorithms to enhance and automate various aspects of binary analysis. For instance, data-driven disassembly and decompilation methods can produce more accurate and human-readable code \cite{hosseini2022beyond,armengol2024slade,hu2024degpt,tan2024llm4decompile}. Deep learning techniques can infer variable types and function signatures, reconstructing higher-level abstractions from binary code \cite{pei2021stateformer,jin2022symlm,chen2022augmenting,jiang2023nova}. Additionally, generative models can generate binary code summaries, providing a concise and high-level description of the code's functionality \cite{al2023extending,xiong2023hext5}. Encoder models are also employed to generate semantic embeddings of binary code, identifying similarities and differences to aid in vulnerability detection and patch analysis \cite{yang2021asteria,yang2023asteria,wang2024clap}. These data-driven approaches herald a future in which binary analysis will be more automated, accurate, and comprehensive, reducing the manual effort required and coping with increasingly complex software systems.

\vspace{-0.9ex}
\subsection{Large Language Models}\label{sec:llm}
In the context of early sequential language tasks, encompassing both natural and programming languages, task-specific model fine-tuning has shown promising performance \cite{dou2023towards}. Fine-tuning involves updating the model weights and enhancing its performance on specific tasks by learning the relationship between the input and output of a specific downstream task dataset.

As a phenomenon-level technology, the emergence and rapid development of large language models has triggered disruptive changes in related fields. For example, ChatGPT \cite{chatgpt}, LLama \cite{Llama2023llama}, Claude \cite{claude}, etc., which usually contain billions or even hundreds of billions of parameters, have been trained on massive text data, and have powerful language understanding and generation capabilities. Since LLMs encapsulate comprehensive knowledge, they can be applied to downstream tasks using a novel method called in-context learning \cite{dai2022can}, eliminating the need for extensive downstream datasets for fine-tuning. In-context learning allows the model to perform specific tasks directly by providing task-related contextual information without updating its parameters \cite{brown2020language}.

In this paper, we adopt the method of in-context learning to guide LLMs to understand the binary analysis tasks from multiple perspectives, thus facilitating a comprehensively evaluation of LLMs' performance on binary analysis.

\vspace{-0.5ex}
\section{Motivation}\label{sec:moti}
\vspace{-0.5ex}
In this section, we discuss the motivation by first outlining the challenges faced in evaluating LLMs on binary analysis tasks in \S\ref{sec:challenge}, then elucidating our motivation through a case story of a real-world analysis scenario in \S\ref{sec:case}, and finally detailing our solutions in \S\ref{sec:solutions}.

\vspace{-0.8ex}
\subsection{Challenges}\label{sec:challenge}
\vspace{-0.3ex}
Recently, numerous LLMs have been deployed to tackle software engineering problems, where they have demonstrated superior capabilities in a wide range of tasks such as program understanding \cite{nam2024using}, vulnerability detection \cite{gao2023far}, and automatic program synthesis \cite{wei2023copiloting,jiangICSE2023}, increasing developer productivity and streamlining all aspects of software development \cite{hou2024large}.

Correspondingly, benchmarks for various related tasks have emerged \cite{guo2024codeeditorbench,li2024deveval}, such as HumanEval \cite{chen2021evaluating} for code generation and Defect4j \cite{just2014defects4j} for automated program repair. These benchmarks are critical as they standardize the evaluation process and provide researchers with clear metrics and datasets, which systematically evaluate the effectiveness of different models and create a competitive environment to accelerate technological advancement. However, in the domain of binary analysis, no official benchmarks have been released yet. And implementing such a benchmark is far from straightforward due to the following challenges:

\vspace{0.5ex}
\noindent\textbf{Lack of Reliable Data Sources and Standard Preprocessing Criteria.} The source and quality of binaries significantly impact the validity and reliability of benchmarks. Currently, there is a lack of standardized data collection frameworks to ensure dataset quality and consistency. Additionally, the absence of a unified preprocessing process for binary files, including compilation environment settings, disassembly or decompilation tool selection, and ground-truth identification, hinders meaningful comparisons between models. The black-box nature of LLMs exacerbates this issue. Ensuring that evaluation data is not included in the training set of LLMs, thereby preventing data leakage from compromising the credibility of benchmark results, is a key consideration.

\vspace{0.5ex}
\noindent\textbf{Multiple Tasks of Binary Analysis Process.} Existing works on binary analysis primarily concentrate on isolated tasks, such as binary code similarity detection \cite{wang2024clap}, decompilation \cite{tan2024llm4decompile} or binary code summarization \cite{jin2023binary}. However, binary analysis is inherently a multifaceted process that requires extracting diverse information types from binaries and understanding the complex dependencies between them \cite{david2020neural}. If benchmark that focus solely on individual task, it fails to capture the integrated capabilities required for effective analysis, thus limiting the comprehensiveness and applicability of the evaluation and failing to represent the true performance of LLMs in real-world applications.

\vspace{0.5ex}
\noindent\textbf{Complexity of Realistic Binary Analysis Scenarios.} The benchmark should avoid simplifying the complexity and diversity encountered in real-world binary analysis, e.g., avoid relying on single data sources, which fail to capture the wide range of project types and contexts in actual engineering environments. And this narrow data scope does not adequately represent the diverse challenges analysts face when working on binaries from different industries, technologies, and applications. Second, the benchmark should pay attention to the actual workflow of reverse engineers in binary analysis, thereby fully reflecting the multifaceted and complex challenges encountered in practice.
\begin{wrapfigure}{r}{7.5cm}
	\centering
        \scalebox{1.00}{
	\includegraphics[width=1.00\linewidth]{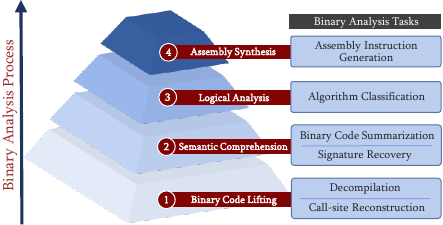}
        }
    \vspace{-3ex}
	\caption{Binary analysis process in case story in \S\ref{sec:case}.}
    \label{fig:case}
\end{wrapfigure}

\vspace{-2.2ex}
\subsection{Insights}\label{sec:case}
\vspace{-0.5ex}
In this section, we present a case story in a real-world binary analysis scenario and then motivate the ideas behind \sysname by describing the details of the work of reverse engineers. 

As shown in Figure \ref{fig:case}, consider a scenario where an enterprise's security team discovers a piece of suspected malware that is spreading across the corporate intranet and appears to steal data and compromise systems. The team is tasked with quickly understanding how this software works to develop effective defenses.

\Circled{1} \textbf{Binary Code Lifting}: The security team immediately begins preliminary analysis of the captured binary executable file, using reverse engineering tools to reconstruct the function call relationships. This helps them understand the software's basic control flow and module dependencies, laying the foundation for subsequent in-depth analysis. As the analysis deepens, they use decompilation technology to convert the binary code into a high-level programming language. This allows for a more intuitive view of the code logic and structure, greatly improving understanding of the malicious software's behavior.

\Circled{2} \textbf{Semantic Comprehension}: During decompilation, the team realizes the need for more contextual information to understand the code's functionality. They work on restoring key information such as function and argument names, argument types, and function return types. By restoring the signature, code readability is greatly improved, allowing for quick identification of potentially malicious code. Due to time constraints, to quickly grasp the full picture of the code, the team generated concise and informative code summaries for each function. These summaries help them understand the overall functionality and behavior of the software without having to examine the code line by line, significantly improving efficiency.

\Circled{3} \textbf{Logical Analysis}: They also discover several complex algorithm codes. To determine their specific purpose and functionality, they categorize and identify these code snippets, helping the team understand the encryption algorithms, compression algorithms, or other key technologies used within the software. 

\Circled{4} \textbf{Assembly Synthesis}: In the final analysis stage, the team generates assembly instructions for specific functionalities to test and validate their defense measures, simulating and reproducing the malware's behavior. This enables the team to verify their defense strategy's effectiveness and further optimize security measures.

Through this series of binary analysis tasks, the security team systematically and comprehensively analyzes the malware's working mechanism. They identify malicious behavior patterns, develop effective defense strategies, and successfully combat threats.

\vspace{-0.7ex}
\subsection{Solutions}\label{sec:solutions}
To address the challenges outlined in \S\ref{sec:challenge}, combined with the analysis of the case story in \S\ref{sec:case}, we propose the following solutions:

\vspace{0.5ex}
\noindent\textbf{Establish Data Collection and Preprocessing Criteria.} Enhancing the quality of benchmarks for binary analysis hinges on establishing robust criteria for data collection and preprocessing. In this regard, inspired by related works \cite{gendercare,li2024deveval,sawada2023arb,zeng2024coderujb} and following standards \cite{NIST,international2011systems,heimann2014ieee}, we advocate that data sources should cover 5 dimensions of realisticity, diversity, quality, credibility, and maintenance, aiming to ensure that the dataset meets the actual requirements of the real-world. Furthermore, we emphasize standardized preprocessing processes, which require defining protocols for compilation, disassembly, and decompilation as well as metadata extraction, ground-truth identification, data filtering, and leak checking. We aim to minimize biases and discrepancies, thereby bolstering the reliability and validity of benchmark assessments.

\vspace{0.5ex}
\noindent\textbf{Enable Multifaceted Task Assessment.} To truly gauge the performance of LLMs in binary analysis, benchmarks should assess a range of interconnected tasks in the binary analysis lifecycle, reflecting the complex nature of real-world analysis workflows. Specifically, our benchmark includes six comprehensive tasks such as decompilation, function signature recovery, and assembly instruction generation. These multifaceted assessments will more comprehensively and accurately reflect the capabilities of LLMs.

\vspace{0.5ex}
\noindent\textbf{Simulate Real-World Analysis Complexity.} In this paper, we seek to replicate the complex challenges encountered in real-world binary analysis scenarios. To start, We amalgamate data sources across different project domains to ensure comprehensive coverage of topics to accommodate the diversity inherent in real-world environments. Subsequently, as described in \S\ref{sec:case}, we dissect the complex challenges faced by reverse engineers in practical binary analysis efforts. In response, we crafted a series of assessment tasks to reflect the complexity of real-world analysis.

\section{\sysname Benchmark}\label{sec:bench}
\begin{figure*}[t]
	\centering
    \scalebox{1.02}{
	   \includegraphics[width=1\linewidth]{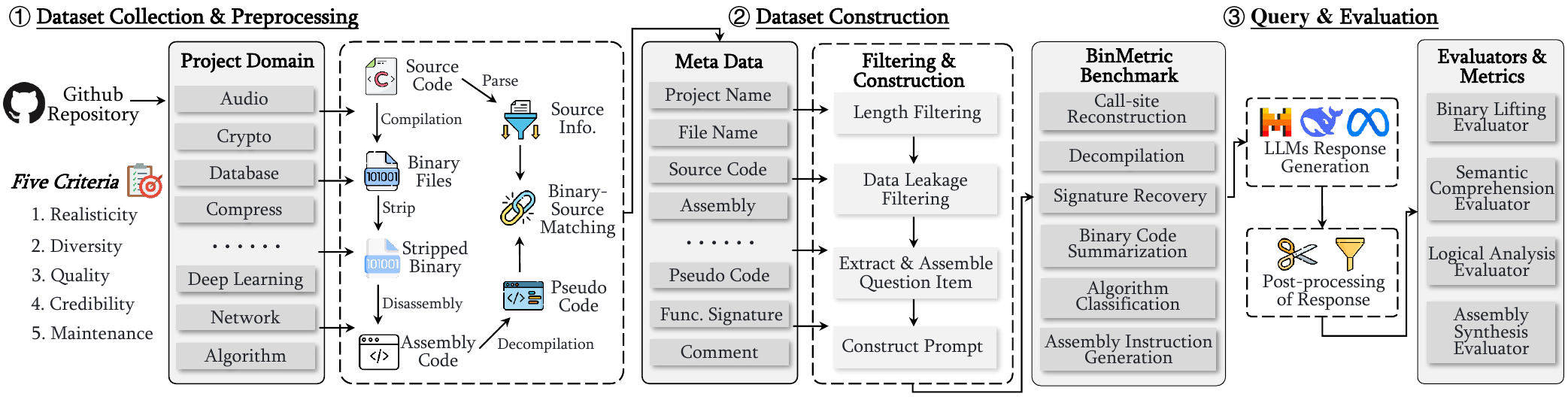}
    }
    \vspace{-4.0ex}
	\caption{Overview framework of \sysname benchmark.}
    \vspace{-2.1ex}
    \label{fig:overview}
\end{figure*}

\vspace{-0.6ex}
This section provides an overview of the binary analysis tasks included in \sysname (detailed in \S\ref{sec:task}) and outlines the dataset collection, preprocessing, and construction in \S\ref{sec:datacollection} and \S\ref{sec:datacon}. The overview framework of \sysname is shown in Figure~\ref{fig:overview}.

\vspace{-0.8ex}
\subsection{Binary Analysis Tasks}\label{sec:task}
\vspace{-0.6ex}
\sysname contains six representative binary analysis tasks that reflect the challenges faced by human reverse engineers in the binary analysis scenario, and LLMs also confront similar obstacles.

\vspace{0.2ex}
\noindent \textbf{Call-site Reconstruction (CSR):}
Function call relationships reveal the basic control flow and dependencies between modules in a program. Call-site reconstruction is a critical step in reverse engineering, with the primary goal of identifying and reconstructing function calls from the provided assembly code, including function name and its calling parameters. We define the input of the task as an assembly code piece and the function call location specified therein, and the output is the corresponding representation of the function call in the high-level source code. The task is evaluated by comparing the textual consistency between the output and the original source code, ensuring that the recovered call sites accurately reflect the intent and structure of the original code.

\vspace{0.2ex}
\noindent\textbf{Decompilation (DEC):}
Decompilation is essential for an intuitive and in-depth understanding of binary programs. The main goal of this task is to reconstruct a human-readable high-level programming language representation, such as C or C++, based on assembly code \cite{armengol2024slade,jiang2023nova,tan2024llm4decompile}. We define the input of this task as an assembly code with function granularity, while the output is the corresponding high-level language code. This task is evaluated using the CodeBLEU \cite{ren2020codebleu} metric, which comprehensively considers the syntax correctness and semantic accuracy of the generated decompiled code by comparing it with the original source code.

\vspace{0.2ex}
\noindent\textbf{Signature Recovery (SR):} Function signature reflects the interface information of the function, including function name, parameter type \& name, and return type, which is crucial for understanding the functionality and behavior of the program \cite{pei2021stateformer,chen2022augmenting,jin2022symlm}. We define the input of this task as decompiled pseudo code from stripped binaries, and the output is the complete function signature. This task uses text consistency metrics to evaluate the matching between the recovered signature and the function signature in the original source code.

\vspace{0.2ex}
\noindent\textbf{Binary Code Summarization (BCS):}
Code summaries can help understand the core functions and operations of a program without delving into complex code details. The main objective of this task is to extract key information from binary code and generate a natural language summary that concisely reflects its essence and functionality \cite{al2023extending,xiong2023hext5}. We define the input of this task as decompiled pseudo code, and the output is its corresponding natural language summary. This task uses text consistency metrics for evaluation to ensure that prediction is consistent with ground truth in terms of relevance and clarity.

\vspace{0.2ex}
\noindent\textbf{Algorithm Classification (AC):}
This task involves identifying and classifying algorithms or algorithmic patterns within binary code, thereby revealing key operations of the program such as sorting, encryption, etc., which are crucial in understanding the function and purpose of the code segment during security analysis. We define the input of this task as decompiled pseudo code and the output as its corresponding algorithm category label. This task is evaluated via accuracy metric. 

\vspace{0.2ex}
\noindent\textbf{Assembly Instruction Generation (AIG):}
Analyzing malware and fixing program vulnerabilities without access to the source code often require modification of assembly code, \emph{i.e}, the ability of assembly synthesis. 
The input of the AIG task is a natural language description of specific functionality, while the output is an assembly instruction fragment corresponding to the functionality. 
In order to echo whether the generated assembly instruction can accurately implement the intended function, we use syntax correctness and execution correctness to assess it. Furthermore, we also employ text consistency to measure the reliability of the generation. 

\vspace{-0.7ex}
\subsection{Data Collection and Preprocessing}\label{sec:datacollection}
\vspace{-0.3ex}
\subsubsection{\textbf{Data Collection.}}  
Creating a reliable benchmark dataset is crucial for evaluating the binary analysis capabilities of LLMs. To ensure the robustness of the dataset and reflect real-world scenarios, we have set the following criteria for our code sources:

\setlength{\leftmargini}{9pt}
\begin{itemize}
    \vspace{-0.2ex}
	\item \textbf{Realisticity}: the code should be derived from complete, human-developer-written real-world projects to ensure relevance and practicality. It should not contain toy programs, incomplete or synthesized snippets.
	\item \textbf{Diversity}: the code should encompass a broad range of fields and application domains, ensuring that the dataset is not biased towards specific types of applications. This variety helps in assessing the models' performance across different contexts. It can be quantified by the domains covered by the projects.
	\item \textbf{Quality}: high-quality code is essential, characterized by clear structure, logical organization, reasonable naming conventions, etc., which help reduce potential ambiguity or errors in the analysis process. It can be quantified by average number of Stars and Forks of project on Github.
    \item \textbf{Credibility}: ideally, the code should come from projects maintained by well-known or reputable developers or organizations. This enhances the reliability and applicability of the dataset in real-world scenarios.
    \item \textbf{Maintenance}: the code from regularly updated, actively maintained, and well-documented projects is preferred because it reflects current programming practices and standards. It can be quantified by the average number of Releases and Commits of projects.
\end{itemize}

\begin{wraptable}{r}{7.6cm}
    \centering
    \caption{Data sources of our benchmark dataset.}
    \vspace{-0.7ex}
    \setlength{\tabcolsep}{0.99mm}
    \scalebox{0.8}{
    \begin{tabular}{l@{\hspace{10pt}}c@{\hspace{6pt}}!{\vline}!{\vline}@{\hspace{6pt}}l@{\hspace{5pt}}c}
        \toprule
            \textbf{Project} & \textbf{Domain} & \textbf{Project} & \textbf{Domain} \\
        \midrule
            audio \cite{aubio} & Audio & Llama2.c \cite{Llama2.c} & Deep Learning \\
            miniaudio \cite{miniaubio} & Audio & Whisper.cpp \cite{whisper.cpp} & Deep Learning \\
            OpenSSL \cite{OpenSSL} & Crypto & Mongoose \cite{Mongoose} & Web  \\
            libsodium \cite{libsodium} & Crypto & libhv \cite{libhv} & Web  \\
            Redis \cite{Redis} & Database  & Curl \cite{Curl} & Network  \\
            SQLite \cite{SQLite} & Database & Masscan \cite{Masscan} & Network  \\
            ImageMagick \cite{ImageMagick} & Image & Libexpat \cite{Libexpat} & Format \\
            Libvips \cite{Libvips} & Image & Ultrajson \cite{Ultrajson} & Format \\
            7z \cite{7z} & Compress & FFmpeg \cite{FFmpeg} & Video \\
            zstd \cite{zstd} & Compress & C-Algorithms \cite{TheAlgorithms} & Algorithm \\
        \bottomrule
    \end{tabular} }
    \vspace{-0.5ex}
    \label{tab:dataset_construction} 
\end{wraptable}

These criteria are inspired by many existing works and follow some industry standards. The National Institute of Standards and Technology (NIST) has proposed criteria for trustworthy AI \cite{NIST}, while international standards and best practices related to software lifecycle processes and quality assurance processes, such as ISO/IEC 25010:2011 \cite{international2011systems}, IEEE Std-730-2014 \cite{heimann2014ieee}, et.al, have put forward requirements for attributes such as quality, credibility, and maintenance. Many LLM benchmarks \cite{gendercare,li2024deveval,sawada2023arb,zeng2024coderujb} emphasize attributes such as realisticity, diversity, quality, and credibility.

Following these criteria, as shown in Table~\ref{tab:dataset_construction}, we curate a diverse selection of 20 real-world C language projects from GitHub, focusing on those with the highest star ratings to ensure high credibility, excellent code quality, and maintenance standards. These projects have an average of 18.48K Stars and 3.9K Forks, 75.8 Releases and 15.5K Commits, spanning eleven domains, including audio, image, web, crypto, network, algorithm, and more, ensuring real coding practice and diversity.

\vspace{-0.2ex}
\subsubsection{\textbf{Data Preprocessing}}  \hspace{0pt}

\noindent As shown in Figure~\ref{fig:overview}, our preprocessing pipeline involves the following steps:

\vspace{0.1ex}
\noindent\textbf{Compile, Strip and Decompile.}
Initially, we compile the projects selected on an Ubuntu 22.04 OS, targeting the x86-64 architecture. During compilation, we include DWARF \cite{Dwarfformat} debugging information to ensure detailed metadata is available for subsequent alignment. Each project uses its default compiler settings, which encompass various compilers and optimization levels, to produce binaries that reflect typical compilation environments. After compilation, we employ the \texttt{strip} command to remove all symbolic information from the binaries, mimicking the conditions encountered in real-world binary analysis where symbolic information is often unavailable. Subsequently, we utilize IDA Pro \cite{IDA} to disassemble and decompile the stripped binaries, obtaining assembly instruction sequences and decompiled pseudo code.

\vspace{0.1ex}
\noindent\textbf{Source Code Information Extraction.} 
We use the srcML \cite{Maletic2015Exploration} tool to analyze and parse the source files, extracting key information including function signatures, function implementations, human-written summaries, etc. The srcML converts the source files into XML format, allowing us to accurately extract and process this information through XML parsing technology. This extracted information will be structurally stored for subsequent alignment. 

\vspace{0.1ex}
\noindent\textbf{Binary-Source Alignment.} We use DWARF debugging information to align source code and binary code, which can record the functions, variables in the binary code, and their locations (including source file name, line number, and column number) in the source code. In this way, we can accurately match the assembly instruction sequence and decompiled code with the corresponding location in the source code. 

\vspace{-0.5ex}
\subsection{Data Construction}\label{sec:datacon}
\vspace{-0.1ex}
\subsubsection{\textbf{Data Filtering}}  
Considering that extremely long code snippets may exceed the input limit of LLMs, resulting in incomplete analysis, and very short code snippets often lack the necessary semantic richness to provide meaningful context for evaluation, we perform double threshold filtering to remove code snippets that are too short or exceed the context window limit of LLMs, ensuring the feasibility of our evaluation.

Furthermore, ensuring that our benchmark dataset is not included in the training set of LLMs is crucial for the effectiveness of evaluation. All our evaluation data are disassembled or decompiled codes, which come from binary files compiled by ourselves and stripped of symbol information. This has greatly prevented our dataset from being collected into training sets by LLMs' developers. To further verify this, we employ the Google search engine to check if any of the code appears in clear text on the internet. Any disassembled or decompiled code found through an exact whole-word searches is removed from the dataset, which strengthens the credibility of our dataset.

\vspace{-0.3ex}
\subsubsection{\textbf{Extract and Assemble Question Item}}  \hspace{0pt}
After data preprocessing and filtering, we obtain high-quality alignment metadata from both binary and source code. These metadata sets form the foundation for constructing the question items of \sysname.

For the DEC and SR tasks, we randomly sample 250 pairs of input pseudo code and their output ground-truth from the metadata sets. For the CSR task, we randomly sample 70 input assembly code snippets, and each snippets is manually annotated with the call site locations to be recovered and the corresponding ground-truth.

For the BCS task, we initially consider using human-written comments extracted from source files as labels, similar to previous work \cite{al2023extending, xiong2023hext5}. However, we observe that less than 20\% of the functions have comments written by developers. Worse yet, not all comments provide functional summaries and often include noisy content with significant variations in quality and style. Consequently, using human-written comments as ground-truth is deemed unreliable. Inspired by recent studies \cite{dagdelen2024structured, tan2024large} that utilize LLMs like ChatGPT to perform data annotation tasks with reasonable reliability, we leverage ChatGPT to generate summaries of the source code, outlining the function's purpose and functionality. We select a total of 250 pseudo code-summary pairs, and each generated summary undergoes manual review for correctness, ensuring high-quality ground truth of the BCS task.

For the AC task, we sample from the C-Algorithms \cite{TheAlgorithms} project in our dataset. This project includes various algorithm implementations in C language, and we select 80 pseudo-code snippets and manually annotate the respective algorithm category. For the AIG task, we carefully design 100 clear and accurate instructions to construct prompts, aiming to guide the LLMs to generate assembly code snippets that follow Intel syntax for given functionalities, such as implementing the bubble sort algorithm. We also equip it with a set of program test cases containing inputs and corresponding outputs to verify the functional correctness of the generated assembly code. 

Overall, the above tasks, which required manual annotation, review, and verification altogether, cost about 60 man-hours.

\vspace{-0.5ex}
\section{Implementation Design}\label{sec:detail}
\vspace{-0.1ex}
\subsection{Large Language Models Setup}
\vspace{-0.1ex}
We evaluate a total of 12 large language models from 5 different model families, including a variety of locally deployable open-source models, and API-callable closed-source models. 

These LLMs are selected to meet the following criteria:

\textbf{(1) State-of-the-Art:} they are widely regarded as the most advanced and capable LLMs, ranking at the top of various leaderboards, such as EvalPlus \cite{EvalPlus} and LLM-Perf \cite{llmperf}.

\textbf{(2) Extensive Code Pre-training:} these models have been pre-trained on large datasets that include substantial amounts of code, enhancing their ability to comprehend and utilize programming languages effectively.

\textbf{(3) Text and Code Generation Abilities:} the models demonstrate strong capabilities not only in natural language generation but also in generating and understanding code, making them versatile for both text and code-related tasks.

\textbf{(4) Instruction Following Proficiency:} we choose the instruction-tuned version to ensure that the model is good at following detailed instructions, which is a critical skill for accurately performing complex binary analysis tasks.

\vspace{-0.5ex}
\subsubsection{\textbf{Open-source Large Language Models}} 

We locally deploy 10 open-source LLMs from 4 different model families, including:

\textbf{Llama2 \cite{Llama2023llama} / CodeLlama \cite{codellama2023rozière}:}
The Llama2 family of LLMs is developed and publicly released by Meta. It is a series of pretrained and fine-tuned generative text models ranging in scale from 7 billion to 70 billion parameters. We choose its \texttt{Llama-2-7b-chat-hf} version. CodeLlama is a code-specialized version of Llama2, designed for general code synthesis and understanding. We deploy the \texttt{CodeLlama-7b-Instruct-hf} and \texttt{CodeLlama-34b-Instruct-hf} versions.

\textbf{DeepSeek \cite{bi2024deepseek} / DeepSeek-Coder \cite{deepseekcoder}:}
DeepSeek LLM is developed by DeepSeek-AI, and it has been trained from scratch on a vast dataset of 2 trillion English and Chinese tokens. We use its \texttt{deepseek-llm-7b-chat} version fine-tuned on extra instruction data. DeepSeek-Coder is a series of code language models, which achieveing excellent performance among open-source code LLMs on multiple programming languages and various benchmarks. We deploy its \texttt{deepseek-coder-7b\\-instruct-v1.5} and \texttt{deepseek-coder-33b-instruct} versions.

\textbf{WizardCoder \cite{wizardcoder2023luo}:}
WizardCoder is an advanced code generation model developed by the WizardLM team, which empowers Code LLMs with complex instruction fine-tuning, by adapting the Evol-Instruct method to the domain of code. For experiments, we deploy the \texttt{WizardCoder-15B-V1.0} and \texttt{WizardCoder-33B-V1.1} versions.

\textbf{Mistral \cite{mixtral2024jiang}:}
Mistral is publicly released by Mistral-AI team, and we focus on its \texttt{Mistral-7B\\-Instruct-v0.2} version, which has been instruction-tuned and equips with multiple advanced attention mechanisms to improve inference speed and reduce inference cost. We also deploy its variant, \texttt{Mixtral-8x7B-Instruct-v0.1}, a Sparse Mixture of Experts (SMoE) generative model.

\vspace{-0.3ex}
\subsubsection{\textbf{Closed-source Large Language Models}} \hspace{0pt}
\vspace{-0.1ex}

\textbf{GPT-3.5 \cite{chatgpt} / GPT-4 \cite{openai2024gpt4}:} Developed by OpenAI, GPT-3.5 and GPT-4 are among the most advanced and widely-used large language models. These models are accessed via OpenAI's API and are known for their exceptional performance across a broad spectrum of nature language and programe language tasks. Their ability to handle complex prompts and generate high-quality responses makes them ideal for evaluating binary analysis tasks.

\begin{table*}
    \centering
    \caption{Base prompt templates of LLMs. \textcolor{customgreen}{\{Example\}} represents the golden demonstration example in the One-shot prompts, and \textcolor{customred}{\{··· ···\}} represents the specific input for each piece of data.}
    \vspace{-0.5ex}
    \small
    \renewcommand{\arraystretch}{1.18}
    \setlength{\tabcolsep}{0.6mm}{
    \scalebox{0.67}{
    \begin{tabular}{c|m{18.8cm}}
        \hline
        \multirow{1}{*}{\textbf{Tasks}} & \multirow{1}{17cm}{\centering \textbf{Prompt Template}} \\
        \hline \hline
        \makecell{Call-site \\ Reconstruction} & Please imagine you are an experienced binary reverse engineer. The following is a disassembled assembly code, your task is to understand its semantics and behavior, and output call point in the form of C source code at \textcolor{customred}{'call sub\_5F57E'}, including only a descriptive function name and function parameters, wrapped in three backticks (\textasciigrave\textasciigrave\textasciigrave), do not explain. \textcolor{customgreen}{\{Example\}}. Input assembly function:   \textasciigrave\textasciigrave\textasciigrave \textcolor{customred}{\{Assmebly Code\}} \textasciigrave\textasciigrave\textasciigrave \\
        \hline
        Decompilation &  Please imagine you are an experienced binary reverse engineer. The following is a disassembled assembly code, your task is to understand it and output its corresponding C source code, wrapped in three backticks (\textasciigrave\textasciigrave\textasciigrave), do not explain. \textcolor{customgreen}{\{Example\}}. Input assembly function:   \textasciigrave\textasciigrave\textasciigrave \textcolor{customred}{\{Pseudo Code\}} \textasciigrave\textasciigrave\textasciigrave  \\ 
        \hline
        \makecell{Signature \\ Recovery} &  Please imagine you are an experienced binary reverse engineer. The following is a stripped decompiled C function, your task is to understand it and output the descriptive function signature in its corresponding source code. This includes the function name, parameter list and its type, and return value type. Wrap the output with three backticks (\textasciigrave\textasciigrave\textasciigrave), do not explain. \textcolor{customgreen}{\{Example\}}. Input decompiled C function:   \textasciigrave\textasciigrave\textasciigrave \textcolor{customred}{\{Pseudo Code\}} \textasciigrave\textasciigrave\textasciigrave  \\ 
        \hline
        \makecell{Binary Code \\ Summarization} & Please imagine you are an experienced binary reverse engineer. The following is a stripped decompiled C function, your task is to understand it and generate a short comment to the function describing its functionality.  \textcolor{customgreen}{\{Example\}}. Input decompiled C function:   \textasciigrave\textasciigrave\textasciigrave \textcolor{customred}{\{Pseudo Code\}} \textasciigrave\textasciigrave\textasciigrave    \\
        \hline
        \makecell{Algorithm \\ Classification} & Please imagine you are an experienced binary reverse engineer. The following is a stripped decompiled C function, your task is to understand it and output its algorithm class from the following list: [Sorting, Searching, Numerical Methods, Hash, Conversions, Math, Dynamic Programming, Cipher, ... ...]. Wrap the output with three backticks (\textasciigrave\textasciigrave\textasciigrave), do not explain. \textcolor{customgreen}{\{Example\}}. Input decompiled C function:   \textasciigrave\textasciigrave\textasciigrave \textcolor{customred}{\{Pseudo Code\}} \textasciigrave\textasciigrave\textasciigrave\\
        \hline
        \makecell{Assembly \\ Instruction \\ Generation}  & \textcolor{customgreen}{\{Example\}}. Design an x64 architecture assembly code for \textcolor{customred}{\{a bubble sort algorithm, which requires an array to be input from the terminal, and then the terminal outputs the sorted result \}}. The generated assembly code is required to be wrapped in three backticks (\textasciigrave\textasciigrave\textasciigrave), can be compiled into an executable program by gcc, and does not contain comments.\\
        \hline
        \end{tabular} }}
    \label{tab:prompt_detail}
    \vspace{-1.5ex}
\end{table*}

\vspace{-0.3ex}
\subsection{Base Prompt Templates} \label{sec:prompt}
\vspace{-0.3ex}
To fully utilize the in-context learning ability of LLMs, we choose the one-shot prompt strategy. Meanwhile, we also analysis the performance of zero-shot prompt in \S\ref{sec:otherfactors}. Although zero-shot prompt performs well in some natural language tasks, it often lacks sufficient contextual information when dealing with complex binary analysis tasks, resulting in low accuracy and consistency of model output. While the chain-of-thought method can guide the model to reason step by step, it has various limitations in practical applications, including the model's context window length, reasoning efficiency, and uncertain performance improvements.

The detailed one-shot prompts for different tasks we designed are displayed in Table~\ref{tab:prompt_detail}. We use carefully selected golden demonstration example as part of the prompts to help the model understand the task context and expected output format. Moreover, we use role-play prompts \cite{kong2023better, chen2023unleashing} to position the model as an "experienced binary reverse engineer" to more effectively convey the task requirements and reduce ambiguity in the model's understanding of the task. We enclose the code in the prompt with three backticks (\textasciigrave\textasciigrave\textasciigrave) to clearly describe the code format. We also require the model output to be wrapped in three backticks to facilitate parsing during post-processing, thereby reducing the interference of noisy text.

\vspace{-0.8ex}
\subsection{Evaluators and Metrics}
In this section, we discuss the evaluators and metrics used to assess the performance of LLMs on binary analysis tasks. Our evaluation framework is meticulously designed and consists of 4 evaluators: binary lifting evaluator, semantic comprehension evaluator, logical analysis evaluator, and assembly synthesis evaluator, each of which aims to measure a specific aspect of binary analysis capabilities.

\vspace{0.2ex}
\noindent\textbf{\Circled{1} Binary Lifting Evaluator.} The primary purpose of this evaluator is to assess the performance of LLMs to convert binary code in assembly form into a higher-level representation, which is crucial for reconstructing the structure and behavior of binary programs. It is applicable to both CSR and DEC tasks, and the evaluation metrics involved are Rouge-L \cite{lin-2004-rouge} and CodeBLEU \cite{ren2020codebleu} respectively. Rouge-L measures the textual consistency between the generated call-site information 
 and the reference, while CodeBLEU, as an metric for commonly used code synthesis tasks, is used to evaluate the syntactic and semantic logic similarity of the decompiled code to the corresponding source code.

\vspace{0.2ex}
\noindent\textbf{\Circled{2} Semantic Comprehension Evaluator.} To measure the depth of LLMs’ understanding and interpretation of binary code, and their ability to capture the latent meaning and intent behind code snippets, we designed this evaluator for SR and BCS tasks. By using BLEU-1 \cite{papineni-etal-2002-bleu}, METHOR \cite{meteor2009Lavie}, and Rouge-L \cite{lin-2004-rouge} together, we provide a comprehensive assessment that covers precise word matching, semantic flexibility and richness, and structural coherence.

\vspace{0.2ex}
\noindent\textbf{\Circled{3} Logical Analysis Evaluator.} Accurate logic analysis is essential to understand the specific behavior and purpose of a code snippet. This evaluator assesses the capability of LLMs to comprehend the logical structure of algorithmic snippets in binary code, and to determine whether they can identify and classify specific algorithms or algorithmic patterns. It is suitable for AC tasks, with the metric being Accuracy, which measures the correctness of LLMs in algorithm classification.

\vspace{0.2ex}
\noindent\textbf{\Circled{4} Assembly Synthesis Evaluator.} This evaluator is designed for the AIG task to assess LLMs' ability to generate accurate, executable assembly instruction snippets from natural language descriptions. The evaluation involves three metrics: Syntactic correctness, Execution correctness, and Rouge-L \cite{lin-2004-rouge}. Syntax correctness is determined by checking if the generated assembly code can be compiled without errors, ensuring it adheres to proper syntax rules. Since executability is one of the most crucial features of a program, we verify execution correctness using pre-designed test cases to confirm that the code performs the intended functionality. To avoid the security risks brought by directly executing the generated code, we use Docker to create an isolated environment to safely execute the assembly code. And Rouge-L intuitively measures the textual consistency of the generated code and the expected output.

\vspace{-0.8ex}
\section{Evaluation}\label{sec:eval}

\vspace{-0.2ex}
\subsection{Research Questions (RQs)}
Our empirical study aims to investigate the following research questions to explore the binary analysis capabilities of LLMs:

\vspace{-0.5ex}
\setlength{\leftmargini}{9pt}
\begin{itemize}
\item \textbf{RQ1: What is the overall effectiveness of LLMs in binary analysis?} This research question aims to evaluate the general capabilities of LLMs, understand how LLMs handle the inherent complexity and challenges of analyzing binary code, and provide a benchmark for their overall effectiveness in this domain.
\item \textbf{RQ2: Which LLM we investigated performs the best? And which type of LLMs performs better?} We specifically focus on performance comparisons between open-source and closed-source LLMs, general and code-specific LLMs, and LLMs of varying parameter sizes, with the goal of understanding the strengths and limitations of each LLM in detail. 

\item \textbf{RQ3: How efficient are LLMs? And what factors affect the effectiveness of LLMs?} This research question investigates the efficiency of LLMs, as well as the impact of prompt design and code length on the effectiveness of LLMs. Our goal is to provide comprehensive guidelines for optimizing the use of LLMs in binary analysis applications.
\end{itemize}

\vspace{-0.5ex}
\subsection{Experiment Setup}
\subsubsection{\textbf{Environments}}
Our experimental environment operates on an Ubuntu 22.04 server, equipped with two 28-core Intel Xeon 6330 CPUs, 512GB RAM, 64TB storage, and eight NVIDIA RTX A6000 GPUs, each with 48GB of VRAM. These GPUs run Nvidia driver version 525.116.03 along with CUDA version 12.0. 

\vspace{-0.5ex}
\subsubsection{\textbf{Implementation Details}}
The \sysname benchmark and all experiments are conducted using Python 3.8 with PyTorch \cite{PyTorch} 2.0.1, Transformers \cite{Transformers} 4.37.2 and DeepSpeed \cite{DeepSpeed} 0.13.0 packages. For closed-source LLMs, i.e. ChatGPT and GPT4, we call OpenAI’s API to access \texttt{gpt-3.5-turbo-16k-0613} and \texttt{gpt-4-0613} backend models. We download all the open-source LLMs from Huggingface \cite{HuggingFace} and enable half-precision in FP16 for efficient inference. Given the context window limitations of LLMs, we set \texttt{max\_length} to 8192 and \texttt{max\_new\_tokens} to 2048. Considering that when using language models to handle code tasks, most scenarios need to ensure the accuracy of the model responses rather than diversity, so we set the sampling \texttt{temperature} to 0.1, which controls the diversity and creativity of the model response. We also set both \texttt{top\_k} and \texttt{top\_p} to 1 during inference to obtain highly deterministic responses.

\vspace{-0.5ex}
\subsubsection{\textbf{Baseline Methods}}
To better analyze the effectiveness of LLMs, we endeavor to include as many baseline methods as possible for comparison.
For the static analysis tool IDA Pro \cite{IDA}, we utilize its Hex-rays decompiler for CSR, DEC, and SR tasks. However, it does not support the other tasks. For DEC task, we evaluate three versions (1.3B, 6.7B and 33B) of the existing work LLM4Decompile \cite{tan2024llm4decompile}, which is a LLM fine-tuned on the DeepSeek-Coder for decompilation task. 
For BCS task, the only existing open-source methods are BinT5 \cite{al2023extending} and HexT5 \cite{xiong2023hext5}, and we reproduced them. For the other two tasks, i.e., AC and AIG, there are currently no comparable baseline methods, and our \sysname can serve as a potential baseline for future research in these areas.

\vspace{-0.5ex}
\subsection{RQ1: Overall Effectiveness}
\vspace{-0.1ex}

In Table~\ref{tab:effective}, we present the overall effectiveness of various LLMs across different binary analysis tasks. We use \textcolor{customred}{$\uparrow$} and \textcolor{customgreen}{$\downarrow$} to denote the increase and decrease in performance between two rows, respectively. The best scores in each task are highlighted in bold. 

\begin{table}[t]
    \caption{Overall effectiveness of LLMs and baseline method on \sysname. \textcolor{customred}{$\uparrow\!$} and \textcolor{customgreen}{$\downarrow\!$} represent a relative increase and decrease between two rows.}
    \vspace{-1.0ex}
    \renewcommand{\arraystretch}{1.03}
    \setlength{\tabcolsep}{0.83mm}{
    \scalebox{0.6}{
    \begin{tabular}{|c@{\hspace{0pt}}c@{\hspace{0pt}}c|l|l|l|lll|l|lll|}
        \hline
            &  &  & \multicolumn{10}{c|}{\textbf{\sysname Benchmark}}  \\ 
        \cline{4-13}    
            &  &  & \multicolumn{1}{c|}{\textbf{CSR}}  & \multicolumn{1}{c|}{\textbf{DEC}} & \multicolumn{1}{c|}{\textbf{SR}} & \multicolumn{3}{c|}{\textbf{BCS}} & \multicolumn{1}{c|}{\textbf{AC}}  & \multicolumn{3}{c|}{\textbf{AIG}}     \\ 
        \cline{4-13}    
            \multirow{-3}{*}{\textbf{Type}} & \multirow{-3}{*}{\textbf{Model}} & \multirow{-3}{*}{\textbf{Size}}  & \multicolumn{1}{c|}{\textbf{$Rouge$-$L$}}  & \multicolumn{1}{c|}{\textbf{$CodeBLEU$}}  & \multicolumn{1}{c|}{\textbf{$Rouge$-$L$}} & \multicolumn1{c}{\textbf{$BLEU$}} & \multicolumn{1}{c}{\textbf{$METEOR$}} & \multicolumn{1}{c|}{\textbf{$Rouge$-$L$}}  & \multicolumn{1}{c|}{\textbf{$Acc.$}} & \multicolumn{1}{c}{\textbf{\makecell{$Syntax$ \\[-0.5ex] $Corr.$}}}  & \multicolumn{1}{c}{\textbf{\makecell{$Execution$ \\[-0.5ex] $Corr.$}}} & \multicolumn{1}{c|}{\textbf{$Rouge$-$L$}}   \\ 
        \hline\hline
            \multicolumn{1}{|c}{\multirow{6}{*}{\textbf{\makecell{Baseline \\ Methods}}}} & \multirow{3}{*}{\textbf{\makecell{LLM4-\\Decompile}\cite{tan2024llm4decompile}}} & \textbf{1.3B} & - & 21.53 & - & - & - & - & - & - & - & - \\  
            & & \textbf{6.7B} & - & 22.97 \begingroup\scriptsize\textcolor{customred}{\smash{\raisebox{0.25cm}{($\uparrow\!\!6.7\%$)}}}\endgroup & - & - & - & - & - & - & - & - \\
            & & \textbf{33B} & - & 24.48 \begingroup\scriptsize\textcolor{customred}{\smash{\raisebox{0.25cm}{($\uparrow\!\!6.6\%$)}}}\endgroup & - & - & - & - & - & - & - & - \\
        \cline{2-13} 
            & \multirow{1}{*}{\textbf{IDA Pro \cite{IDA}}} & \textbf{-} & 8.52 & 18.56 & 10.27 & - & - & - & - & - & - & - \\
            & \multirow{1}{*}{\textbf{BinT5 \cite{al2023extending}}} & \textbf{220M} & - & - & - & 31.58 & 1.82 & 4.28 & - & - & - & - \\
            & \multirow{1}{*}{\textbf{HexT5 \cite{xiong2023hext5}}} & \textbf{223M} & - & - & - & 36.92 & 4.99 & 7.65 & - & - & - & - \\
        \hline
            & \multirow{1}{*}{\textbf{Llama2}} & \textbf{7B} & 3.83 & 24.71 & 6.51 & 35.83 & 21.64 & 19.00 & 15.00 & 1.00 & 0.00 & \textbf{31.58} \\
        \cdashline{2-3}
            & & \textbf{7B} & 4.91 \begingroup\scriptsize\textcolor{customred}{\smash{\raisebox{0.25cm}{($\uparrow\!\!28.2\%$)}}}\endgroup & 23.62\begingroup\scriptsize\textcolor{customgreen}{\smash{\raisebox{0.25cm}{($\downarrow\!\!4.4\%$)}}}\endgroup & 14.49\begingroup\scriptsize\textcolor{customred}{\smash{\raisebox{0.25cm}{($\uparrow\!\!122.6\%$)}}}\endgroup & 16.32\begingroup\scriptsize\textcolor{customgreen}{\smash{\raisebox{0.25cm}{($\downarrow\!\!54.5\%$)}}}\endgroup  & 16.81\begingroup\scriptsize\textcolor{customgreen}{\smash{\raisebox{0.25cm}{($\downarrow\!\!22.3\%$)}}}\endgroup & 13.05\begingroup\scriptsize\textcolor{customgreen}{\smash{\raisebox{0.25cm}{($\downarrow\!\!31.3\%$)}}}\endgroup & 36.25\begingroup\scriptsize\textcolor{customred}{\smash{\raisebox{0.25cm}{($\uparrow\!\!141.7\%$)}}}\endgroup & \textbf{68.00}\begingroup\scriptsize\textcolor{customred}{\smash{\raisebox{0.25cm}{($\uparrow\!\!6700\%$)}}}\endgroup & 0.00\begingroup\scriptsize\textcolor{customred}{\smash{\raisebox{0.25cm}{($\uparrow\!\!0.0\%$)}}}\endgroup & 24.44\begingroup\scriptsize\textcolor{customgreen}{\smash{\raisebox{0.25cm}{($\downarrow\!\!22.6\%$)}}}\endgroup \\
            & \multirow{-2}{*}{\textbf{CodeLlama}} & \textbf{34B} & 6.17 \begingroup\scriptsize\textcolor{customred}{\smash{\raisebox{0.25cm}{($\uparrow\!\!25.7\%$)}}}\endgroup & 20.60\begingroup\scriptsize\textcolor{customgreen}{\smash{\raisebox{0.25cm}{($\downarrow\!\!12.8\%$)}}}\endgroup & 26.59\begingroup\scriptsize\textcolor{customred}{\smash{\raisebox{0.25cm}{($\uparrow\!\!83.5\%$)}}}\endgroup & 29.04\begingroup\scriptsize\textcolor{customred}{\smash{\raisebox{0.25cm}{($\uparrow\!\!77.9\%$)}}}\endgroup  & 28.89\begingroup\scriptsize\textcolor{customred}{\smash{\raisebox{0.25cm}{($\uparrow\!\!71.9\%$)}}}\endgroup & 22.89\begingroup\scriptsize\textcolor{customred}{\smash{\raisebox{0.25cm}{($\uparrow\!\!75.4\%$)}}}\endgroup & 65.00\begingroup\scriptsize\textcolor{customred}{\smash{\raisebox{0.25cm}{($\uparrow\!\!79.3\%$)}}}\endgroup & 63.00\begingroup\scriptsize\textcolor{customgreen}{\smash{\raisebox{0.25cm}{($\downarrow\!\!7.4\%$)}}}\endgroup & 2.00\begingroup\scriptsize\textcolor{customred}{\smash{\raisebox{0.25cm}{($\uparrow\!\!inf\%$)}}}\endgroup & 23.98\begingroup\scriptsize\textcolor{customgreen}{\smash{\raisebox{0.25cm}{($\downarrow\!\!1.9\%$)}}}\endgroup \\
        \cline{2-13} 
            & \multirow{1}{*}{\textbf{DeepSeek}} & \textbf{7B} & 4.92 & 16.63 & 8.95 & 40.44 & 26.97 & 23.92 & 20.00 & 17.00 & 0.00 & 14.50 \\
        \cdashline{2-3}
            & & \textbf{7B} & 2.99 \begingroup\scriptsize\textcolor{customgreen}{\smash{\raisebox{0.25cm}{($\downarrow\!\!39.2\%$)}}}\endgroup & 21.36\begingroup\scriptsize\textcolor{customred}{\smash{\raisebox{0.25cm}{($\uparrow\!\!28.4\%$)}}}\endgroup & 11.66\begingroup\scriptsize\textcolor{customred}{\smash{\raisebox{0.25cm}{($\uparrow\!\!30.3\%$)}}}\endgroup & 32.45\begingroup\scriptsize\textcolor{customgreen}{\smash{\raisebox{0.25cm}{($\downarrow\!\!19.8\%$)}}}\endgroup  & 30.23\begingroup\scriptsize\textcolor{customred}{\smash{\raisebox{0.25cm}{($\uparrow\!\!12.1\%$)}}}\endgroup & 23.44\begingroup\scriptsize\textcolor{customgreen}{\smash{\raisebox{0.25cm}{($\downarrow\!\!2.0\%$)}}}\endgroup & 47.50\begingroup\scriptsize\textcolor{customred}{\smash{\raisebox{0.25cm}{($\uparrow\!\!137.5\%$)}}}\endgroup & 41.00\begingroup\scriptsize\textcolor{customred}{\smash{\raisebox{0.25cm}{($\uparrow\!\!141.2\%$)}}}\endgroup & \textbf{4.00}\begingroup\scriptsize\textcolor{customred}{\smash{\raisebox{0.25cm}{($\uparrow\!\!inf\%$)}}}\endgroup & 24.54\begingroup\scriptsize\textcolor{customred}{\smash{\raisebox{0.25cm}{($\uparrow\!\!69.2\%$)}}}\endgroup \\
            & \multirow{-2}{*}{\textbf{DeepSeek-Coder}} & \textbf{33B} & 6.35 \begingroup\scriptsize\textcolor{customred}{\smash{\raisebox{0.25cm}{($\uparrow\!\!112.4\%$)}}}\endgroup & 22.77\begingroup\scriptsize\textcolor{customred}{\smash{\raisebox{0.25cm}{($\uparrow\!\!6.6\%$)}}}\endgroup & 19.97\begingroup\scriptsize\textcolor{customred}{\smash{\raisebox{0.25cm}{($\uparrow\!\!71.3\%$)}}}\endgroup & 43.58\begingroup\scriptsize\textcolor{customred}{\smash{\raisebox{0.25cm}{($\uparrow\!\!34.3\%$)}}}\endgroup  & 30.41\begingroup\scriptsize\textcolor{customred}{\smash{\raisebox{0.25cm}{($\uparrow\!\!0.6\%$)}}}\endgroup & 27.60\begingroup\scriptsize\textcolor{customred}{\smash{\raisebox{0.25cm}{($\uparrow\!\!17.7\%$)}}}\endgroup & 71.25\begingroup\scriptsize\textcolor{customred}{\smash{\raisebox{0.25cm}{($\uparrow\!\!50.0\%$)}}}\endgroup & 16.00\begingroup\scriptsize\textcolor{customgreen}{\smash{\raisebox{0.25cm}{($\downarrow\!\!61.0\%$)}}}\endgroup & 1.00\begingroup\scriptsize\textcolor{customgreen}{\smash{\raisebox{0.25cm}{($\downarrow\!\!75.0\%$)}}}\endgroup & 12.64\begingroup\scriptsize\textcolor{customgreen}{\smash{\raisebox{0.25cm}{($\downarrow\!\!48.5\%$)}}}\endgroup \\
        \cline{2-13} 
            & & \textbf{15B} & 0.05 & 22.62 & 27.04 & 25.76 & 27.44 & 20.16 & 51.25 & 0.00 & 0.00 & 4.02 \\
            & \multirow{-2}{*}{\textbf{WizardCoder}} & \textbf{33B} & 8.34 \begingroup\scriptsize\textcolor{customred}{\smash{\raisebox{0.25cm}{($\uparrow\!\!inf\%$)}}}\endgroup & 23.62\begingroup\scriptsize\textcolor{customred}{\smash{\raisebox{0.25cm}{($\uparrow\!\!4.4\%$)}}}\endgroup & \textbf{28.12}\begingroup\scriptsize\textcolor{customred}{\smash{\raisebox{0.25cm}{($\uparrow\!\!4.0\%$)}}}\endgroup & \textbf{45.93}\begingroup\scriptsize\textcolor{customred}{\smash{\raisebox{0.25cm}{($\uparrow\!\!78.3\%$)}}}\endgroup  & 28.09\begingroup\scriptsize\textcolor{customred}{\smash{\raisebox{0.25cm}{($\uparrow\!\!2.4\%$)}}}\endgroup & \textbf{27.93}\begingroup\scriptsize\textcolor{customred}{\smash{\raisebox{0.25cm}{($\uparrow\!\!38.5\%$)}}}\endgroup & 75.00\begingroup\scriptsize\textcolor{customred}{\smash{\raisebox{0.25cm}{($\uparrow\!\!46.3\%$)}}}\endgroup & 1.00\begingroup\scriptsize\textcolor{customred}{\smash{\raisebox{0.25cm}{($\uparrow\!\!inf\%$)}}}\endgroup & 0.00\begingroup\scriptsize\textcolor{customred}{\smash{\raisebox{0.25cm}{($\uparrow\!\!0.0\%$)}}}\endgroup & 9.95\begingroup\scriptsize\textcolor{customred}{\smash{\raisebox{0.25cm}{($\uparrow\!\!147.5\%$)}}}\endgroup \\
        \cline{2-13} 
            & & \textbf{7B} & 4.63 & 19.51 & 19.49 & 41.39 & \textbf{31.00} & 26.63 & 48.75 & 2.00 & 0.00 & 7.51 \\
            \multirow{-10}{*}{\textbf{\makecell{Open \\ Source}}} & \multirow{-2}{*}{\textbf{Mixtral}} & \textbf{8x7B} & 7.27 \begingroup\scriptsize\textcolor{customred}{\smash{\raisebox{0.25cm}{($\uparrow\!\!57.0\%$)}}}\endgroup & 22.15\begingroup\scriptsize\textcolor{customred}{\smash{\raisebox{0.25cm}{($\uparrow\!\!13.5\%$)}}}\endgroup & 21.91\begingroup\scriptsize\textcolor{customred}{\smash{\raisebox{0.25cm}{($\uparrow\!\!12.4\%$)}}}\endgroup & 43.19\begingroup\scriptsize\textcolor{customred}{\smash{\raisebox{0.25cm}{($\uparrow\!\!4.3\%$)}}}\endgroup  & 29.88\begingroup\scriptsize\textcolor{customgreen}{\smash{\raisebox{0.25cm}{($\downarrow\!\!3.6\%$)}}}\endgroup & 26.73\begingroup\scriptsize\textcolor{customred}{\smash{\raisebox{0.25cm}{($\uparrow\!\!0.4\%$)}}}\endgroup & 65.00\begingroup\scriptsize\textcolor{customred}{\smash{\raisebox{0.25cm}{($\uparrow\!\!33.3\%$)}}}\endgroup & 32.00\begingroup\scriptsize\textcolor{customred}{\smash{\raisebox{0.25cm}{($\uparrow\!\!1500\%$)}}}\endgroup & 1.00\begingroup\scriptsize\textcolor{customred}{\smash{\raisebox{0.25cm}{($\uparrow\!\!inf\%$)}}}\endgroup & 27.89\begingroup\scriptsize\textcolor{customred}{\smash{\raisebox{0.25cm}{($\uparrow\!\!271.4\%$)}}}\endgroup \\
        \hline
            & \textbf{GPT-3.5-Turbo} & / & 4.09 & 21.04 & 19.62 & 36.16 & 30.69 & 25.09 & 60.00 & 15.00 & 0.00 & 23.97 \\
            \multirow{-2}{*}{\textbf{\makecell{Close \\ Source}}} & \textbf{GPT-4} & / & \textbf{9.61} & \textbf{25.99} & 20.69 & 44.48 & 24.81 & 23.25 & \textbf{83.75} & 11.00 & 1.00 & 18.62 \\
        \hline
        \hline
            & \textbf{Average} &  & 5.26 & 22.05 & 18.75 & 36.21 & 27.24 & 23.31 & 53.23 & 22.25 & 0.75 & 18.64 \\
        \hline
    \end{tabular} } }
    \label{tab:effective}
    \vspace{-0.2ex}
\end{table}

\vspace{0.3ex}
\noindent\textbf{Results on Binary Lifting.}
For the binary lifting evaluation, we use Rouge-L and CodeBLEU scores to measure the performance of LLMs in call-site reconstruction (CSR) and decompilation (DEC), respectively. The average scores are 5.26\% and 22.05\% on both tasks. \uline{Rebuilding call-sites from binary functions without any other knowledge about the callee themselves is very challenging for the general LLMs.} 
More specifically, although GPT-4 achieves the highest score of 9.61\% in CSR, manual inspection reveals only a few useful reconstructions. And except for GPT-4, all other LLMs scored below the baseline method IDA Pro, which achieved 8.52\%. WizardCoder-15B has the lowest score of 0.05\% in CSR due to its failure to follow the instructions, and we could barely parse prediction from the output. 

Whereas the overall score for lifting the binary function into the source-code level is relatively high, which implies that \uline{LLMs have the potential to perform the semantic mapping between disassembled code and source code.} 
Specifically, in the DEC task, LLMs demonstrate performance close to average, with DeepSeek-7B and GPT-4 recording the highest and lowest scores of 25.99\% and 16.63\%, respectively. It demonstrates that the decompilation task has a similar hardness for non-tailored models and does not show dramatic performance variations depending on the model size and domain. Compared to its base model DeepSeek-Coder, LLM4Decompile demonstrates a slight performance improvement at the same parameter scale.

\vspace{0.3ex}
\noindent\textbf{Results on Semantic Comprehension.}
To evaluate the semantic comprehension of LLMs on binary code, we conducted two experiments, signature recovery (SR) and binary code summarization (BCS), across all models. \uline{For these two semantic understanding tasks, LLMs perform relatively well and mostly outperform the baseline methods.}

In the SR task, the LLMs achieved an average Rouge-L score of 18.75\%, 
and it is WizardCoder-33B that excels most at this task (scored 28.12\%), while Llama2-7B has struggled in it (scored 6.51\%). Furthermore, our experiments have demonstrated that both the size and domain of models significantly affect the performance in the SR task. Specifically, an increase in model size resulted in a 42.8\% improvement across open-source LLMs. And the code-specific LLMs outperformed general LLMs by 76.45\%. 
On the other hand, for the BCS task, we use three metrics (BLEU, METEOR, and Rouge-L) to provide an aggregate and comprehensive assessment. Wizardcoder-33B has the best mean score of 33.3\% on all three metrics. In contrast, CodeLlama-7B lagged in generating natural language summaries, with a score of 22\%. Similarly, the code-specific variant, DeepSeek-Coder-7B, underperformed compared to its general counterpart, DeepSeek-7B, scoring 28.71\% versus 30.44\%. This disparity suggests that additional pre-training on code data may have adversely affected the model’s natural language expression capabilities. 

\vspace{0.3ex}
\noindent\textbf{Results on Logical Analysis.}
We utilized the Algorithm Classification (AC) task to estimate the capability of LLMs on binary code logical analysis, where we prompt the models to identify the class of pseudo code among a given candidate list. \uline{The overall score of LLMs is 53.23\%, indicating that they can understand high-level semantics and support deeper logical reasoning tasks to a certain extent.}
GPT-4 has achieved a significant first position with an accuracy score of 83.75\%. Conversely, we find that the two small-sized general models performed the worst, with Llama2-7B and DeepSeek-7B achieving scores of 15.00\% and 20.00\%, respectively. 
In this task, the code-specific models show an average 139.6\% outperformance than the general ones. Meanwhile, increasing the model size also brings a notable improvement.

\vspace{0.3ex}
\noindent\textbf{Results on Assembly Synthesis.}
Generating assembly code (AIG) on request is a typical evaluation task for assembly synthesis. We prompt the models to generate a fragment of assembly instructions to implement a specific function, adhering to Intel Syntax. Only the generation that compiled successfully is deemed syntactically correct. CodeLlama-7B excels at the AIG task, achieving a Syntax correctness score of 68\%, which is likely benefiting from additional pre-training on assembly code. In contrast, other models struggle to generate syntactically correct assembly code. Even GPT-4, a leader in DEC tasks, scores only 11\%. Furthermore, we devise test cases for IO verification to assess the semantic execution correctness. Notably, all models found it challenging to pass the IO test, with an average pass rate of less than 1\%. 
Meanwhile, we employ Rouge-L to calculate the textual consistency of the generated text with the reference, offering an additional perspective on the evaluation. Llama2-7B records the highest Rouge-L score (31.58\%), with manual identification revealing more similar code-blocks between its prediction and the reference. 
However, models with significantly lower scores, such as WizardCoder-15B (4.02\%) and Mixtral-7B (7.51\%), fail to follow our instructions to generate assembly code. \uline{The results show that current LLMs are not up to the task of reconstructing detailed low-level operations in binaries, possibly due to the inherent complexity of assembly-level instructions and the lack of domain expertise.}

\vspace{0.2ex}
\begin{tcolorbox}[colback=gray!5,
                  colframe=black,
                  arc=0.8mm, auto outer arc,
                  boxrule=1pt,
                  boxsep=-2pt
                 ]
\textbf{Answering RQ1:} The LLMs evaluated show overall potential for binary analysis, yet all perform poorly on CSR and AIG tasks, indicating that current non-expert models do not effectively address these analysis challenges. Additionally, each model exhibits expertise in specific analysis capabilities. For instance, GPT-4 excels in binary lifting and logical analysis, while WizardCoder and CodeLlama excel in semantic comprehension and assembly synthesis, respectively. 
\end{tcolorbox}

\vspace{-0.5ex}
\subsection{RQ2: LLMs Comparison} \label{sec:compare}
\vspace{-0.3ex}

\begin{figure}
    \centering
    \begin{subfigure}[b]{0.45\linewidth}  
        \centering
        \includegraphics[width=\linewidth]{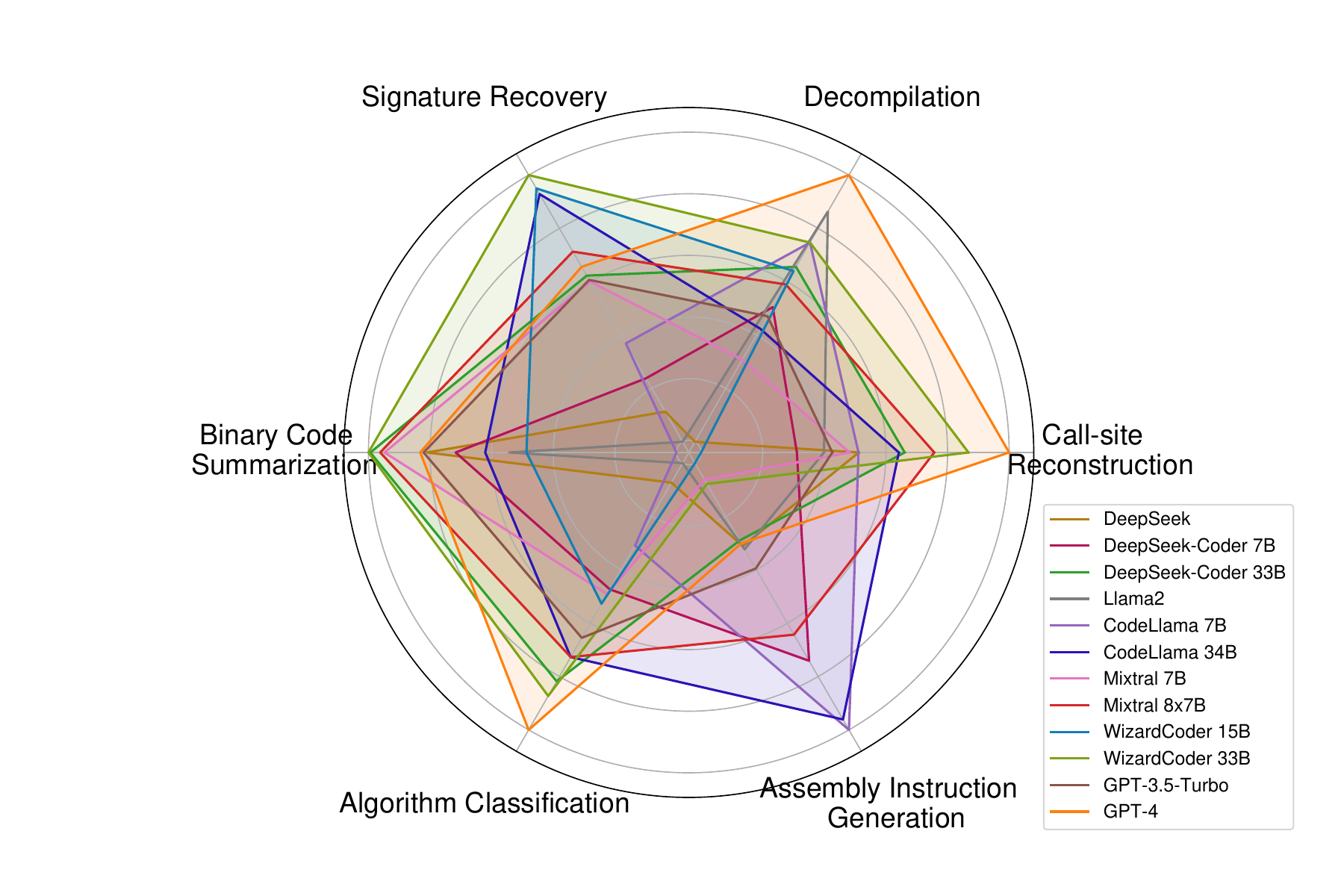}   
        \vspace{-2ex}
        \caption{Relative performance of LLMs on \sysname benchmark against the best in each tasks.}
    \end{subfigure}
    \quad
    \begin{subfigure}[b]{0.51\linewidth}  
        \centering
        \includegraphics[width=\linewidth]{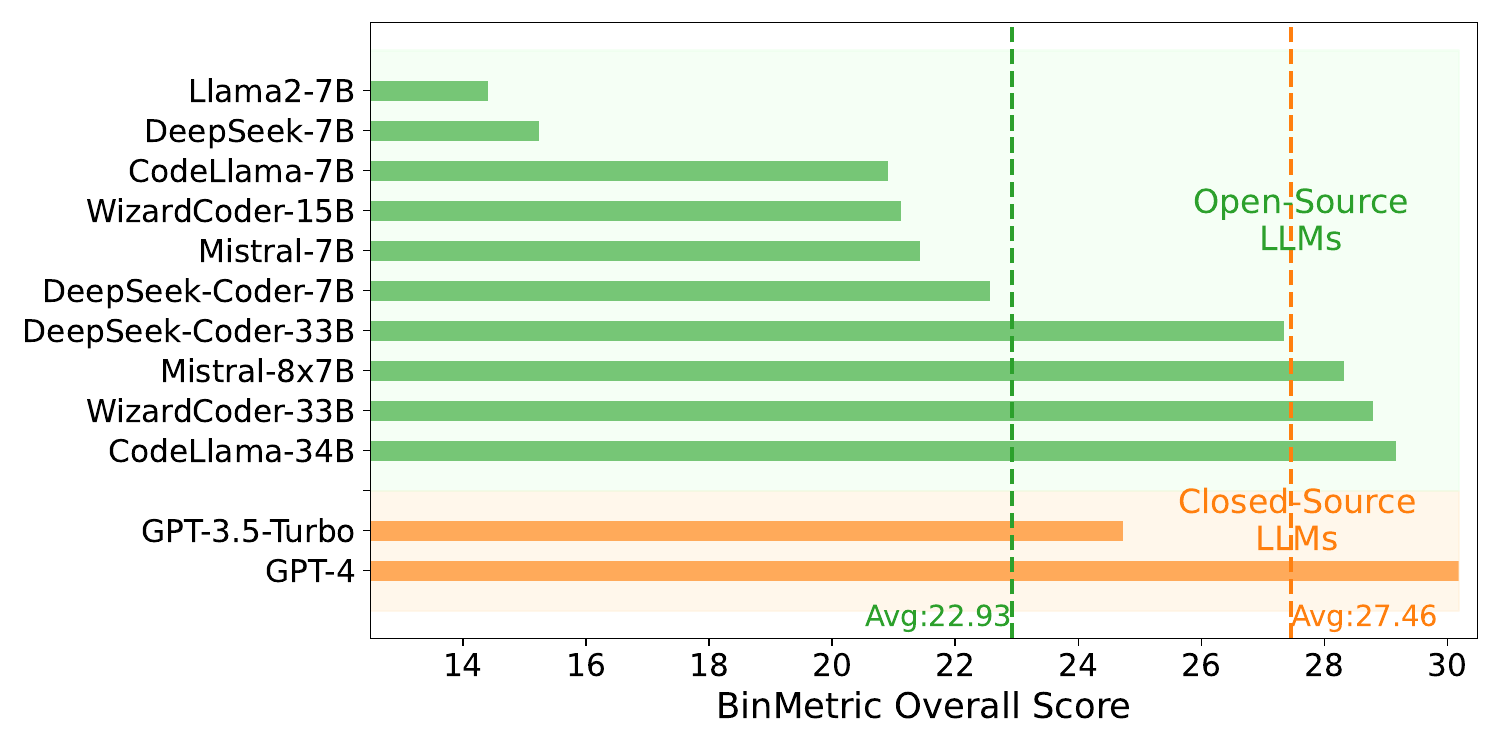}
        \vspace{-2ex}
        \caption{Overall scores of LLMs on \sysname benchmark across 6 tasks. Dashed lines for two LLM types' average.}
    \end{subfigure}
    \vspace{-1ex}
    \caption{Performance comparison between LLMs.}
    \vspace{-2ex}
    \label{fig:rq2}
\end{figure}

As shown in Figure~\ref{fig:rq2} (a), we aggregate the scores of the models across various tasks to construct a radar chart, which illustrates the performance differences among the models. Specifically, we utilize the average score of the metrics used in each task as the task score and calculate the relative scores across the models based on the highest score achieved in each task. In general, GPT-4 dominates the arena, dramatically outperforming the other models in the DEC, CR, and AC tasks, although it underperforms in the AIG task. Trailing closely, WizardCoder-33B, CodeLlama-34B, and Mixtral-8x7B exhibit comparable overall performance, each revealing specific strengths and weaknesses. For instance, WizardCoder-33B, akin to GPT-4, struggles in assembly synthesis, whereas CodeLlama-34B excels in AIG and SR tasks. 

\begin{wrapfigure}{r}{6.5cm}
	\centering
        \scalebox{1}{
	\includegraphics[width=1.00\linewidth]{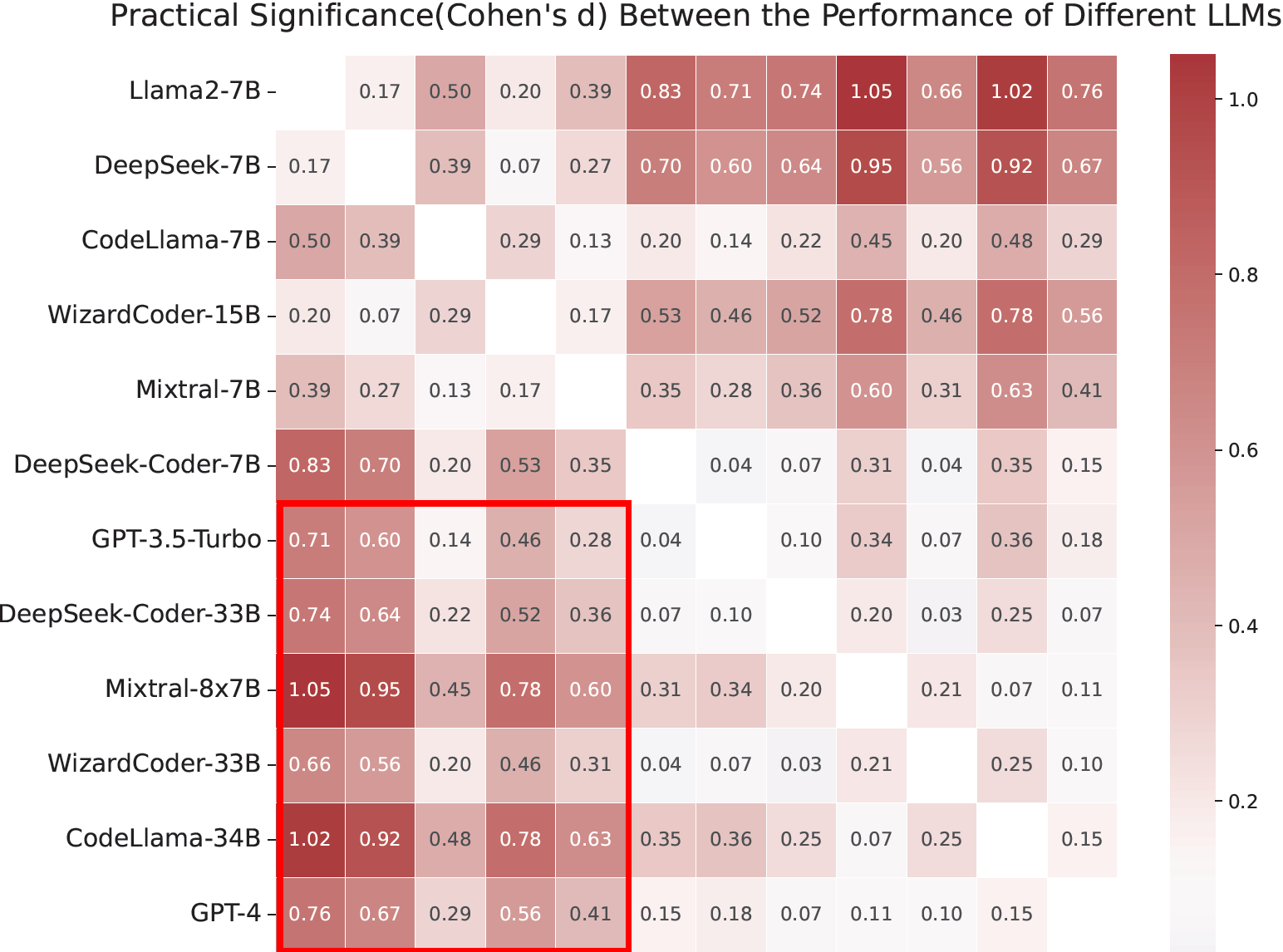}
        }
    \vspace{-4.2ex}
	\caption{Results of practical significance (Cohen's d).}
    \label{fig:hypotest}
\end{wrapfigure}

In addition, we perform statistical hypothesis tests to further analyze the performance differences among LLMs. We use Cohen's d indicator for practical significance analysis, and results are shown in Figure \ref{fig:hypotest}.

\vspace{0.2ex}
\noindent\textbf{Open-source \emph{v.s.} Closed-source LLMs.}
Figure~\ref{fig:rq2} (b) shows the average scores of each model over multiple tasks, and we use dashed lines to show the mean scores for both the open-source and closed-source models. The mean score of the closed-source model, at 27.46\%, is significantly higher than the 22.93\% achieved by its counterparts. However, some open-source models, such as CodeLlama-34B, surpass the average performance of their closed-source ones. This observation encourages the notion that open-source models are not significantly outpaced by closed-source models in binary analysis scenarios, offering strong confidence in developing expert LLMs in the binary domain based on these models.

\vspace{0.1ex}
\noindent\textbf{Parameter Size \& Code-specific Knowledge.}
Our study introduces two common sets of model sizes (7-15B \& 33-8x7B) and involves models with code domain knowledge from multiple LLM families. 
As shown in Figure~\ref{fig:rq2} (b), the models with more parameters generally exhibit superior performance, potentially due to enhanced ability to follow instructions and more effectively embedded information, which can also be observed by the red box in Figure \ref{fig:hypotest}, where they have higher practical significance. Furthermore, domain-specific knowledge within the code domain yields a performance advantage on these models, despite the code dataset seldom includes decompiled binary code.

\vspace{-0.3ex}
\begin{tcolorbox}[colback=gray!5,
                  colframe=black,
                  arc=0.8mm, auto outer arc,
                  boxrule=1pt,
                  boxsep=-2pt
                 ]
\textbf{Answering RQ2:} Based on the evaluation of our benchmark, GPT-4 leads other models in the binary analysis. Meanwhile, we recommend the CodeLlama-34B as the ``open-source winner'', which could be taken as a baseline to facilitate future research. In addition, the results encourage us to use the code-specific model and try larger sizes.
\end{tcolorbox}

\vspace{-0.5ex}
\subsection{RQ3: Efficiency and Other Factors} \label{sec:otherfactors}
\vspace{-0.5ex}
To answer RQ3, we explore the overhead of LLMs on various binary analysis tasks. Meanwhile, We investigate the impact of prompts design and code length on the overall effectiveness of LLMs.

\vspace{0.3ex}
\noindent\textbf{Inference Efficiency.} We measure the inference time of various LLMs across multiple tasks. Considering that closed-source LLMs are affected by network latency, rate limits, etc., our measurements only focus on open-source LLMs. The results are presented in Table~\ref{tab:efficiency}. First, LLMs with larger parameters (33-8x7B) take 1.58$\times$ more time on average for inference compared to smaller LLMs (7-15B). However, as discussed in \S\ref{sec:compare}, larger LLMs generally have stronger analytical capabilities, which requires a trade-off between effectiveness and efficiency. In addition, different tasks show different levels of complexity. For example, the more complex DEC and AIG tasks require an average of 38.17s and 56.12s, respectively, while CSR and SR only require an average of 11.47s and 9.29s.

\begin{wraptable}{r}{8.45cm}
    \caption{Analysis efficiency (in Seconds) of LLMs.}
    \vspace{-0.3ex}
    \renewcommand{\arraystretch}{1.02}
    \setlength{\tabcolsep}{1.1mm}{
    \scalebox{0.8}{
    \begin{tabular}{c@{\hspace{0pt}}crrrrrr|r}
        \hline
            \multicolumn{1}{c}{\multirow{2}{*}{\textbf{Model}}} & \multicolumn{1}{c}{\multirow{2}{*}{\textbf{Size}}}    & \multicolumn{6}{c}{\textbf{\sysname Benchmark}} & \multicolumn{1}{c}{\multirow{2}{*}{\textbf{Avg.}}}  \\ 
        \cline{3-8}    
            &  & \multicolumn{1}{c}{\textbf{CSR}}  & \multicolumn{1}{c}{\textbf{DEC}} & \multicolumn{1}{c}{\textbf{SR}} & \multicolumn{1}{c}{\textbf{BCS}} & \multicolumn{1}{c}{\textbf{AC}}  & \multicolumn{1}{c}{\textbf{AIG}} & \\ 
        \hline\hline
            \multirow{1}{*}{\textbf{Llama2}} & \textbf{7B} & 2.49 & 1.17 & 1.40 & 4.99 &  0.57 & 24.27 & 5.82 \\
            & \textbf{7B} & 27.82 & 23.28 & 2.26 & 42.16 & 58.51 & 58.49 & 35.42\\
            \multirow{-2}{*}{\textbf{CodeLlama}} & \textbf{34B} & 2.58 & 52.24 & 4.51 & 75.03 & 1.71 & 103.96 & 40.01 \\
        \cline{1-9} 
            \multirow{1}{*}{\textbf{DeepSeek}} & \textbf{7B} & 0.66 & 15.01 & 13.21 & 5.13 & 0.43 & 22.54 & 9.50 \\
            & \textbf{7B} & 0.80 & 28.22 & 2.92 & 7.91 & 1.09 & 21.24 & 10.36 \\
            \multirow{-2}{*}{\textbf{DeepSeek-Coder}} & \textbf{33B} & 3.01 & 85.99 & 9.70 & 22.95 & 1.59 & 62.55 & 30.97 \\
        \cline{1-9} 
            & \textbf{15B} & 69.13 & 37.85 & 38.64 & 35.44 & 47.97 & 57.69 & 47.79 \\
            \multirow{-2}{*}{\textbf{WizardCoder}} & \textbf{33B} & 4.12 & 77.98 & 10.63 & 20.15 & 2.12 & 61.22 & 29.37 \\
        \cline{1-9} 
            & \textbf{7B} & 1.03 & 14.42 & 3.86 & 5.95 & 0.44 & 51.55 & 12.88 \\
            \multirow{-2}{*}{\textbf{Mixtral}} & \textbf{8x7B} & 3.07 & 45.59 & 5.72 & 16.59 & 1.09 & 97.73 & 28.30 \\
        \hline \hline
            \textbf{Average} & / & 11.47 & 38.17 & 9.29 & 23.63 & 11.55 & 56.12 & 25.04 \\
        \hline
    \end{tabular} } }
    \label{tab:efficiency}
    \vspace{-0.3ex}
\end{wraptable}

\begin{figure}
    \centering
    \begin{subfigure}[b]{0.47\linewidth}  
        \centering
        \includegraphics[width=\linewidth]{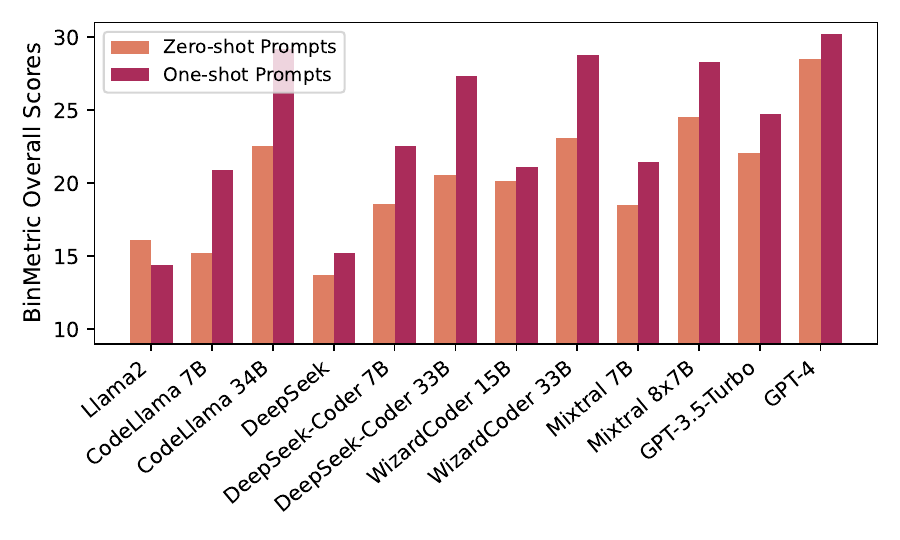}   
        \vspace{-3ex}
        \caption{Overall scores of LLMs on \sysname benchmark with Zero-shot (Left) and One-shot (Right) prompts.}
    \end{subfigure}
    \quad
    \begin{subfigure}[b]{0.47\linewidth}  
        \centering
        \includegraphics[width=\linewidth]{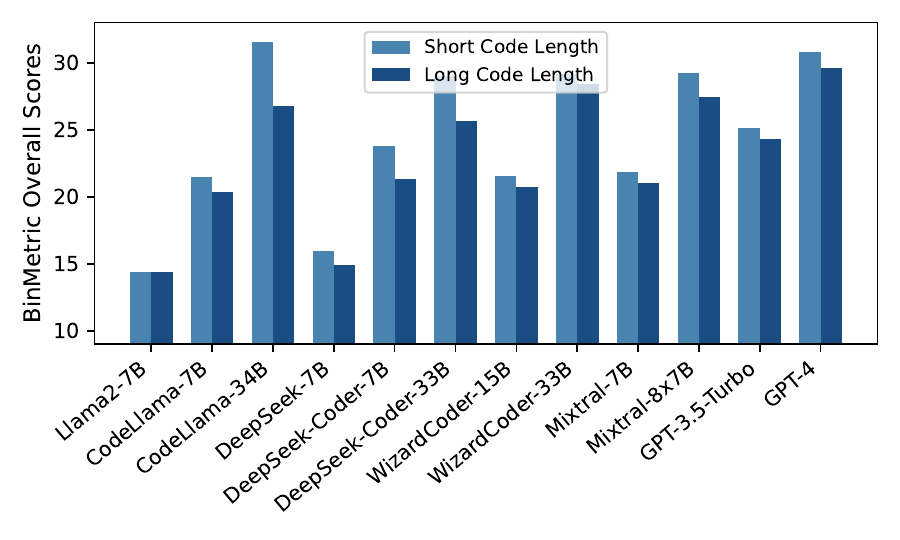}
        \vspace{-3ex}
        \caption{Overall score of LLMs on the \sysname benchmark for short(Left) and long(Right) code input lengths.}
    \end{subfigure}
    \vspace{-2ex}
    \caption{Impact of prompt word and code length on performance}
    \vspace{-2ex}
    \label{fig:rq3}
\end{figure}

\vspace{0.3ex}
\noindent\textbf{Prompts Design.} As discussed in \S\ref{sec:prompt}, the effectiveness of LLMs may be significantly influenced by prompt design. Figure~\ref{fig:rq3} (a) compares the overall scores of LLMs using zero-shot and one-shot prompts. We observe that in most cases, one-shot prompts substantially improved the overall scores of LLMs, with an average increase of 16.65\%. This enhancement underscores the importance of providing demonstration examples to guide the LLMs' understanding and response generation in the complex domain of binary analysis.

However, an exception is noted: the overall score of Llama2-7B decreased by 10.79\%. Upon analyzing its output, we find that it tends to replicate the example output provided in the one-shot prompt rather than generate new responses based on the context. This behavior may be attributed to its inherent capacity limitations, causing the demonstration example to disrupt its performance.

\vspace{0.5ex}
\noindent\textbf{Code Length.}
The input code length is a crucial factor that can significantly influence the effectiveness of LLMs in binary analysis tasks. To investigate this, we divide the problem items of each task into shorter and longer groups according to their code length. Figure~\ref{fig:rq3} (b) compares the overall scores of LLMs on short and long input code. The results indicate a clear trend: the overall scores longer code snippets decrease by an average of 6.00\% compared to shorter ones. This decline is observed across various tasks, highlighting the challenges when handling extended code snippets.

\begin{tcolorbox}[colback=gray!5,
                  colframe=black,
                  arc=0.8mm, auto outer arc,
                  boxrule=0.85pt,
                  boxsep=-2.1pt
                 ]
\textbf{Answering RQ3:} Larger LLMs offer better performance but at the cost of efficiency, with inference time 1.58$\times$ that of smaller LLMs on average. One-shot prompts effectively improve the overall effectiveness of LLMs. Besides, the effectiveness of LLMs is affected by the increase in input code length.
\end{tcolorbox}

\vspace{-0.2ex}
\section{Discussion}\label{sec:discussion}

\vspace{-0.2ex}
\subsection{Potential and Limitations of LLMs for Binary Analysis.}
\vspace{-0.5ex}
In this paper, we conduct an empirical study on multiple LLMs via the \sysname benchmark, revealing their potential and limitations in binary analysis tasks. Specifically, GPT-4 and WizardCoder-34B demonstrate impressive capabilities in algorithm classification and binary code summarization tasks, respectively, indicating that they can accurately extract high-level semantic information of binary codes and support deeper logical reasoning tasks, which are crucial for understanding complex binary files. However, existing LLMs still face challenges in call-site reconstruction and assembly instruction generation tasks. These tasks require the LLMs to reconstruct detailed low-level operations in binaries, which is difficult for current LLMs, probably due to the inherent complexity of assembly-level instructions and the lack of domain expertise.

In summary, LLMs demonstrate both potential and limitations in binary analysis tasks. Future research should focus on addressing these weaknesses, and with the continuous advancement of technology, LLMs still have broad prospects in this field.

\vspace{-0.2ex}
\subsection{Future Development to Binary-Specific LLMs.}
\vspace{-0.5ex}

With the rapid development of foundation LLMs and the emergence of many domain expert LLMs, and given the potential shown by existing LLMs on binary analysis tasks, expert LLMs for binary analysis are feasible and promising. 
Just like source code LLM, which is evaluated in multiple tasks such as code completion, code infilling, and multilingual evaluation, we believe that binary code expert LLM should have comprehensive capabilities to handle multiple complex tasks in the binary analysis lifecycle. As a benchmark leaderboard designed for multi-tasks, \sysname will assess and drive progress in this field.

To achieve this goal, the binary analysis expert LLM should incorporate extensive binary domain knowledge in the pre-training phase, including various assembly languages, compiler optimization modes, instruction set architecture differences, etc., so as to have a deeper grasp of code semantics and structure. The LLM can also be fine-tuned in a targeted manner in conjunction with specific binary analysis tasks to enhance its performance in key tasks. Furthermore, by combining retrieval-augmented generation strategies or designing specialized instruction fine-tuning datasets, the binary domain knowledge of LLM can be further extended. For example, retrieval enhancement strategies can allow LLM to dynamically query external knowledge bases, thereby providing more accurate judgments when processing complex binary patterns. Architectural innovation that supports longer context windows is also an important direction for improving binary expert LLM. Longer binary code snippets can provide more complete context information, thereby improving the analysis accuracy and generation quality of LLM. According to our empirical studies findings, different LLMs exhibit advantages in specific aspects of analysis. Future research could explore integrating the strengths of various LLMs to construct an expert ensemble model for binary analysis.

\vspace{-0.2ex}
\subsection{Threats to Validity.}
\vspace{-0.5ex}

Although we strive to maintain scientific and rigorous when constructing benchmark and conducting empirical studies, there are still some possible limitations that need to be noted and discussed.

\vspace{0.2ex}
\noindent\textbf{Selection of Analysis Tasks.} As we mentioned in Section \ref{sec:case}, we selected tasks that are common in each stage of the binary analysis lifecycle and suitable for solving with LLM. However, it is unrealistic to cover all potential tasks. For example, variable and data structure recovery, we regard them as part of the decompilation task. Although similarity detection is very common in binary analysis, existing solutions almost exclusively use encoder-only models for embedding representation, which is not suitable for sequence-to-sequence LLMs, and small-scale models have achieved excellent performance.

\vspace{0.2ex}
\noindent\textbf{Scale of the Benchmark.} The \sysname benchmark contains 1,000 problem samples across six tasks. While the sample size is sufficient to draw certain conclusions, increasing the size could further stabilize the results. However, expanding the scale is challenging due to the significant manual effort required for annotation. 
For benchmark, we believe that attributes such as being lightweight, easily reproducible, and non-computational resource intensive are of great practical importance, can promote widespread adoption among researchers. Referring to benchmarks in related fields \cite{austin2021program,du2023classeval,yuan2023no,yu2024codereval}, especially HumanEval \cite{chen2021evaluating} for code generation and Defects4J \cite{just2014defects4j} for automated program repair, have sample sizes ranging from 100 to 1,000. Therefore, we consider the current sample size is moderate, with the diversity and representativeness of the data sources ensuring the validity of our evaluation.

\vspace{0.2ex}
\noindent\textbf{Selection of LLMs.} In the current study, our selection of LLMs is limited to a subset of the rapidly developing field of LLMs. Due to computational resource limitations, we are restricted to evaluating open-source models with 7B to 8x7B parameter scales, and some closed-source models. Although these models have been able to reflect the capabilities of LLMs in binary analysis, there is still room for exploration.

\vspace{0.2ex}
\noindent\textbf{Ignorance of obfuscated binaries.} Our \sysname benchmark does not consider any form of binary code obfuscation, which can significantly alter the appearance of disassembled or decompiled code, posing challenges for LLMs in terms of accurate comprehension and interpretation. Therefore, we plan to extend the benchmark to include analysis of obfuscated binary code in future work.

\vspace{-0.2ex}
\section{Conclusion}\label{sec:conclusion}
\vspace{-0.5ex}

In this paper, we conduct a pioneering study on the binary analysis capabilities of LLMs. We design a standardized data collection and preprocessing pipeline, which led to the creation of \sysname, a comprehensive benchmark with 1,000 question items covering 6 representative tasks, reflecting real-world binary analysis challenges. Our empirical study, which test various LLMs, reveals their strengths and limitations on various tasks. The findings indicate that while LLMs show strong potential in binary analysis, significant challenges remain. As LLMs continue to evolve, we believe that \sysname, as a benchmark leaderboard, will become a crucial tool for assessing and driving progress in this field.

\begin{acks}
This work was supported in part by the Natural Science Foundation of China under Grant U20B2047, 62072421, 62002334, 62102386, and 62121002.
\end{acks}

\bibliographystyle{ACM-Reference-Format}
\bibliography{main.bib}

\end{document}